\documentclass[prb,showpacs,twocolumn]{revtex4}
\usepackage{amsmath}
\usepackage{graphicx}
\usepackage{subfigure}
\begin{document}
\title{Magnetic-anisotropy induced spin blockade in a single-molecule transistor}
\author{Guangpu Luo and Kyungwha Park}\email{luogp12@vt.edu}\email{kyungwha@vt.edu}
\affiliation{Department of Physics, Virginia Tech, Blacksburg, Virginia 24061, USA}
\begin{abstract}

We present a new mechanism for a spin blockade effect associated with a change in the type of magnetic anisotropy over
oxidation state in a single molecule transistor, by taking an example of an individual Eu$_{2}$(C$_{8}$H$_{8}$)$_{3}$
molecule weakly coupled to non-magnetic electrodes without linker groups. The molecule switches its magnetization direction
from in-plane to out-of-plane when it is charged. In other words, the magnetic anisotropy of the molecule changes from
easy plane to easy axis when the molecule is charged. By solving the master equation based on model Hamiltonian, we find
that current through the molecule is highly suppressed at low bias independently of gate voltage due to the interplay
between spin selection rules and the change in the type of magnetic anisotropy. Transitions between the lowest magnetic
levels in successive charge states are forbidden because the magnetic levels differ by $|\Delta M| > 1/2$ due to the
change in the type of magnetic anisotropy, although the total spins differ by $|\Delta S|=1/2$. This current suppression
can be lifted by significant ${\mathbf B}$ field, and the threshold ${\mathbf B}$ field varies as a function of the field
direction and the strength of magnetic anisotropy. The spin blockade effect shed light on switching the
magnetization direction by non-spin-polarized current and on exploring effects of this property coupled to other
molecular degrees of freedom.

\end{abstract}
\date{\today}
\pacs{73.23.Hk, 75.50.Xx, 73.63.-b}

\maketitle

\section{Introduction}




Individual nanoscale molecules have been successfully bridged between electrodes within single-molecule transistors or scanning tunneling
microscopy in the laboratory. In particular, studies of electron tunneling through individual {\it anisotropic} magnetic molecules showed
unusual properties ranging from nuclear spin states controlled by electric field \cite{WERN14}, Berry-phase oscillations of
the Kondo peaks \cite{LEUE06}, spin-polarized current induced magnetic switching \cite{MISI07,KHAJ13}, complete current suppression
\cite{HEER06}, large spin filtering \cite{SALV09}, and giant molecular magnetocapacitance \cite{WU13}. In these cases, magnetic anisotropy
induced by spin-orbit coupling and Jahn-Teller distortion plays a crucial role.

\begin{figure*}
\includegraphics[width=0.09\textwidth]{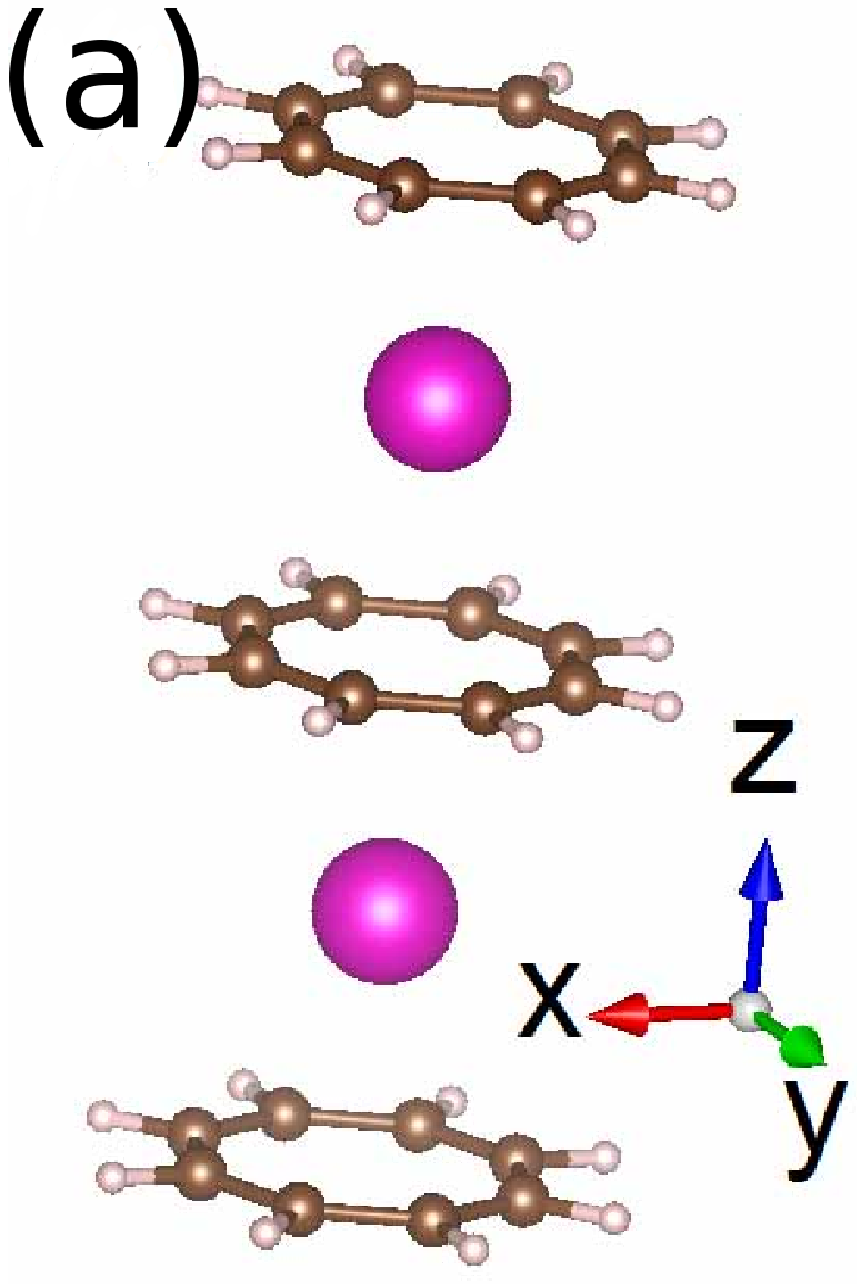}
\hspace{0.2truecm}
\includegraphics[width=0.19\textwidth]{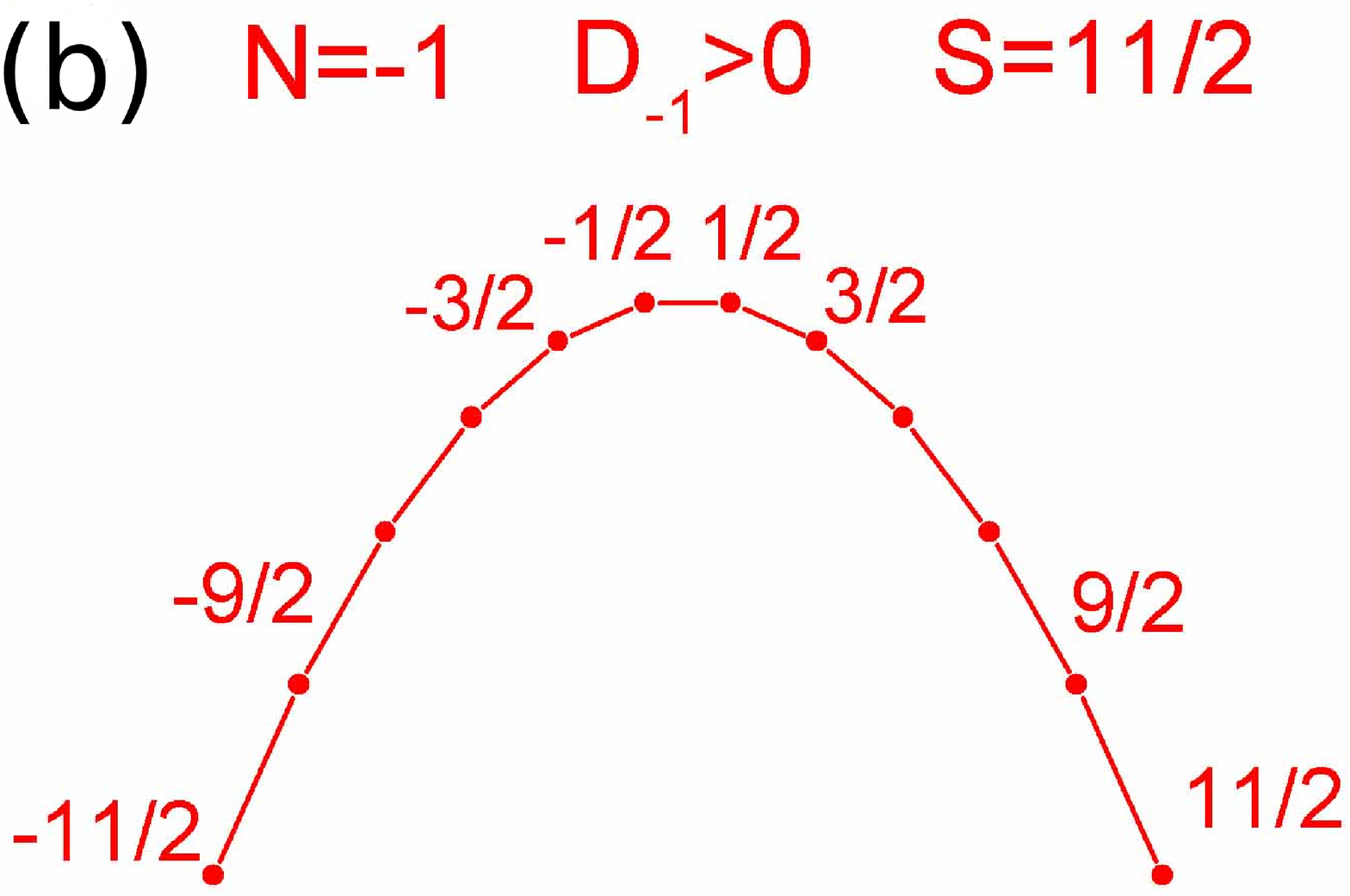}
\includegraphics[width=0.15\textwidth]{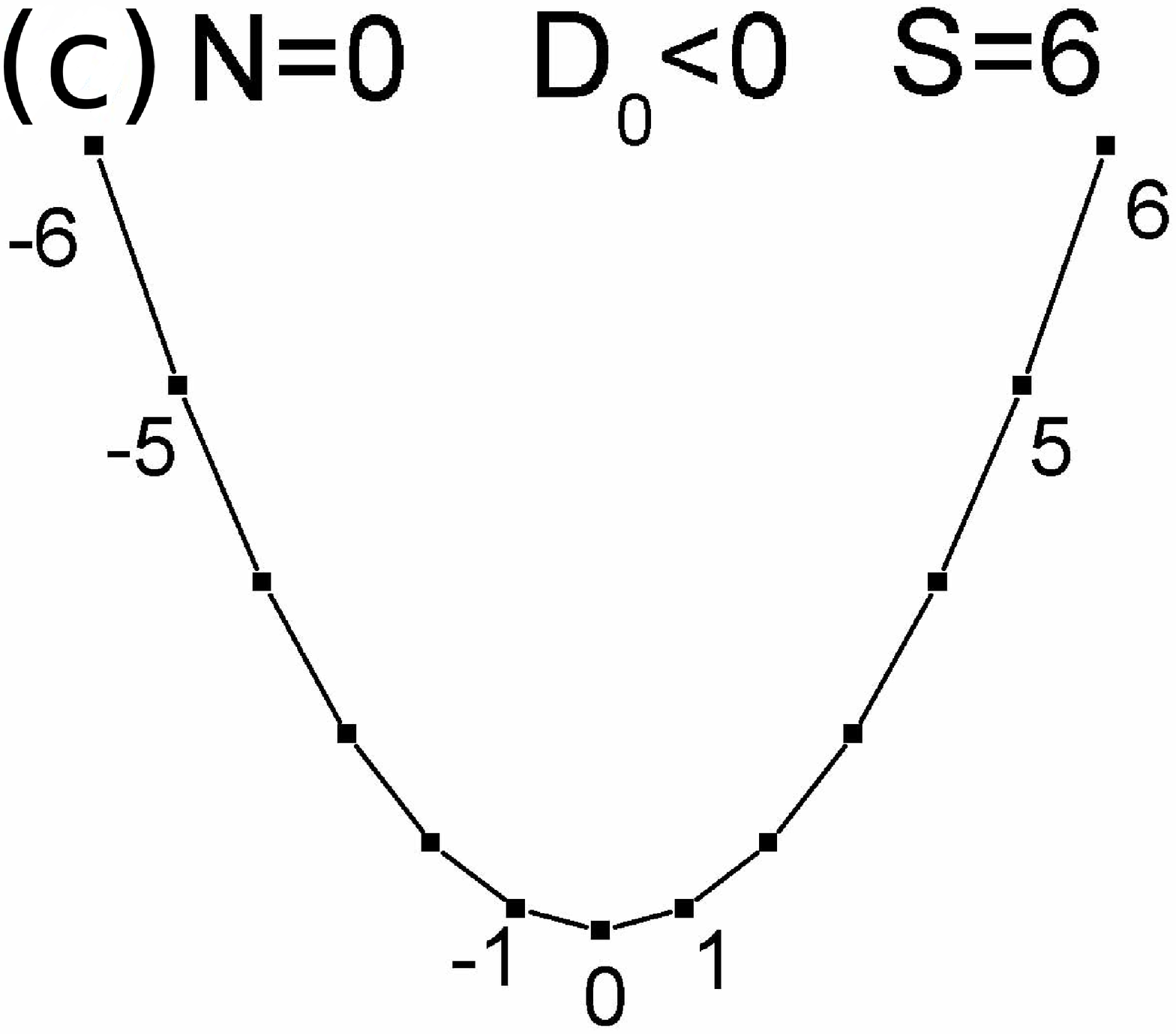}
\hspace{0.2truecm}
\includegraphics[width=0.45\textwidth]{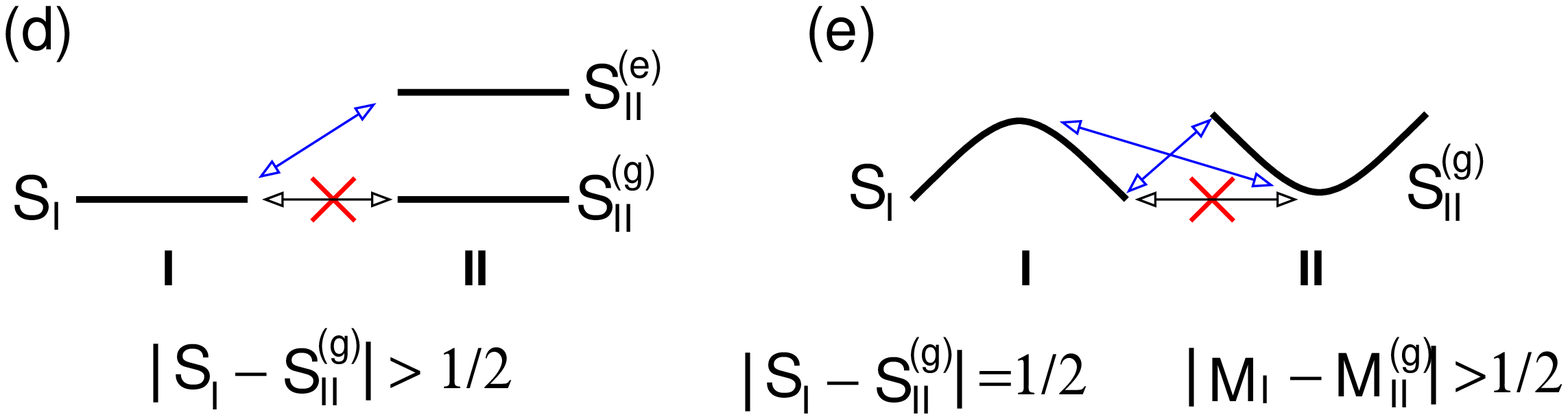}
\caption{(Color online)
(a) Geometry of Eu$_2$(COT)$_3$. (b)-(c) Easy-axis and easy-plane magnetic anisotropy for the cationic ($N=-1$) and neutral ($N=0$) states
of Eu$_2$(COT)$_3$, respectively, where $D_{-1,0}$ are the magnetic anisotropy parameters and $S$ is the total spin of the molecule.
Magnetic levels $M$ are shown for a given total spin $S$, where the $M$ values are eigenvalues of the $z$ component of the total spin
operator. (d) Spin blockade mechanism in Ref.\onlinecite{WEIN95}, where $S_I$, $S_{II}^{(g)}$, and $S_{II}^{(e)}$ are the total spins
of charge state I and ground-state and excited-state of charge state II, respectively. For a given total spin, different $M$
levels are all degenerate due to the absence of magnetic anisotropy. (e) Spin blockade mechanism in this work. In (d) and (e), the total
numbers of electrons in successive charge states I and II differ by unity: $|N_{\mathrm I} - N_{\mathrm {II}}|=1$. Forbidden
transitions are shown with red ``X'' marks, while allowed transitions are shown with blue arrows. For details, see the text.}
\label{fig:geo}
\end{figure*}

Within a magnetic molecule or system, it is difficult to switch the {\it type} of magnetic anisotropy (i.e., easy axis or easy plane)
or the {\it sign} of the magnetic anisotropy parameter
with oxidation state or via charge transfer (from a non-magnetic substrate) because the anisotropy type is often determined by the shape or
orientation of the molecule. Instead, the magnitude of the magnetic anisotropy parameter or the type of magnetic ordering can be relatively
easily varied \cite{HEER06,PARK04,ZYAZ10,ZHAN10,PARK13,LI14}. A change in oxidation state or charge transfer rarely alters orbital character
critical to the anisotropy \cite{ZHAN10,PARK13,LI14}. However, there are some exceptional cases \cite{ATOD08,GAMB09,YANG16}.
First-principles-based calculations revealed that an Eu-sandwiched triple-decker molecule,
Eu$_{2}$(COT)$_{3}$ (COT$=$C$_{8}$H$_{8}$), switches its magnetization direction from
in-plane ($xy$ plane) to out-of-plane (along the $z$ axis) when it is charged \cite{ATOD08} [Fig.~\ref{fig:geo}(a)-(c)].
Fe-based molecules (on copper) \cite{GAMB09} and cobalt films \cite{YANG16}
were shown to change their magnetization directions from in-plane to out-of-plane by oxygen adsorption onto the molecules and by coating
the films with graphene, respectively.

In this work we present a new mechanism for a spin blockade effect associated with a change in the {\it type} of magnetic anisotropy or
in the {\it sign} of the magnetic anisotropy parameter in a single molecule transistor, by taking an example of a single
Eu$_{2}$(COT)$_{3}$ molecule weakly coupled to non-magnetic electrodes. Eu$_{n}$(COT)$_{n+1}$ clusters ($n=1,2,...,18$) have been
synthesized and their magnetic properties have been characterized \cite{MIYA99,HOSO05}. Similar lanthanide-based anisotropic molecules,
[Ln$_n$(COT$^{''}$)$_{n+1}$], were synthesized in crystal phases and they were also formed in solutions \cite{EDEL11,ROY13,TSUJ14}.
By solving the master equation, we find that electron transport via the molecule shows Coulomb blockade-like behavior with one unique
feature: current is highly suppressed at low bias independently of gate voltage. This suppression is caused by the interplay between
spin selection rules and the sign reversal of the magnetic anisotropy parameter. The lowest magnetic levels in successive charge
states differ by greater than 1/2, $|\Delta M| > 1/2$, although the total spins differ by 1/2, due to the sign change in the magnetic
anisotropy parameter [Fig.~\ref{fig:geo}(b), (c), and (e)].
Thus, the spin selection rules prevent transitions between the lowest magnetic levels at zero bias. This spin blockade effect can be
lifted by significant ${\mathbf B}$ field, and the threshold ${\mathbf B}$ field depends on the field direction and the magnetic
anisotropy parameter.

To the best of our knowledge, this kind of a spin blockade effect has not been studied before and it is distinct from other (spin)
blockade effects. A spin blockade effect was reported in quantum dots when the total spins of successive charge states differ by
greater than 1/2 ($|\Delta S| > 1/2$) due to the spin selection rules \cite{WEIN95} [Fig.~\ref{fig:geo}(d)].
Pauli spin blockade was observed when tunneling between singlet
and triplet states is forbidden due to the Pauli exclusion principle in coupled quantum dots \cite{ONO02,JOHN05}. Franck-Condon
blockade occurs when vibron-assisted electron tunneling dominates due to strong electron-vibron coupling \cite{KOCH05,BURZ14}.
The blockade effect observed in Ref.\onlinecite{HEER06} arises from low-lying spin multiplets. Our concept and results are quite general
and so they can be applied to any anisotropic molecules in which the type of magnetic anisotropy can be switched over oxidation state or
charge transfer. The spin blockade effect we report may shed light on controlling the magnetization direction by non-spin-polarized current
without ${\mathbf B}$ field, as well as on exploring effects of this unique property coupled to other molecular degrees of freedom on
electron and thermal transport.

\section{Theoretical Model}


In our model, we assume that an individual Eu$_{2}$(COT)$_{3}$ molecule is bridged between non-magnetic metallic electrodes without linker
groups (i.e., no chemical bonding) and that the molecule can be in a cationic ($N=-1$) or neutral ($N=0$) state by varying gate voltage $V_g$.
We consider only two charge states of the molecule because transport experiments on anisotropic magnetic molecules often reveal only
two charge states \cite{ZYAZ10,BURZ12,JO06} and because the charging energies of anisotropic magnetic molecules are typically several eV
\cite{SALV09,MCCA15}. Figure 3 in Ref.~[\onlinecite{ATOD08}] showed that with spin orbit coupling, the highest occupied molecular orbital
(HOMO) level is split into two singly occupied levels, levels 3 and 4, where level 3 has a higher energy than level 4. We assume that the Eu$_{2}$(COT)$_{3}$ molecule is positively ionized such that the zero-bias Fermi levels of the electrodes are located between level 3 and
level 4. The molecular orbital level for the $N=-1$ ($N=0$) state corresponds to level 4 (level 3). We set the orbital energies of $N=-1$
and $N=0$ as zero and $\varepsilon$, respectively. For $N=-1$, the molecule has the total spin $S=11/2$ with out-of-plane
magnetic anisotropy ($z$ axis) \cite{ATOD08} [Fig.~\ref{fig:geo}(b)]. The $z$ axis is normal to the plane where (COT)$_3$ lie, as shown
in Fig.~\ref{fig:geo}(a). For $N=0$, the total spin becomes $S=6$, and the magnetization direction switches to easy plane ($xy$ plane)
\cite{ATOD08} [Fig.~\ref{fig:geo}(c)]. The molecular Hamiltonian, ${\cal H}_{\rm mol}$, (adapted
from Refs.\onlinecite{MISI07,MISI10,MISI09,KOCH05}) reads
\begin{equation}
\label{eq:H_0}
{\cal H}_{\rm mol}= -D_{N}(S_{z}^{(N)})^2 + (\varepsilon - eV_{g}) \sum_{\sigma}c_{\sigma}^{\dagger}c_{\sigma}
+ g\mu_{B}\textbf{S}^{(N)}\cdot\textbf{B},
\end{equation}
where $D_{N}$ is a magnetic anisotropy parameter for state $N$ with spin $S^{(N)}$. We use $D_{N=-1}=0.0637$~meV for
$S^{(N=-1)}$$=$11/2 and $D_{N=0}$$=$$-$0.0744~meV for $S^{(N=0)}$$=$6 (Ref.~\onlinecite{ATOD08}). We neglect transverse magnetic anisotropy since
it is typically much smaller than the $|D_N|$ value. The second term in ${\cal H}_{\rm mol}$ describes an electron with spin $\sigma$
tunneled to the molecular orbital $\varepsilon$. In our calculations, for convenience, the value of $\varepsilon$ is included in the
value of $V_g$. Here $c_{\sigma}^{\dagger}$ and $c_{\sigma}$
are electron creation and annihilation operators. The last term is the Zeeman energy with $g=2$. Main differences between our model
and model Hamiltonian in Refs.\onlinecite{MISI07,MISI10,MISI09} are as follows: in the latter, (i) three charge states ($N=0, 1, 2$)
were considered with on-site Coulomb repulsion which is needed due to double occupancy at the lowest unoccupied molecular orbital (LUMO);
(ii) at least one of the electrodes was ferromagnetic; (iii) the sign of the magnetic anisotropy parameter remained the same for different
charge states.

The Hamiltonian ${\cal H}_{\rm el}$$=$
$\sum_{\alpha=L,R}\sum_{{\bf k},\sigma}\epsilon_{{\bf k},\sigma}^{\alpha}a_{{\bf k},\sigma}^{\alpha\dagger}a_{{\bf k},\sigma}^{\alpha}$
is for the non-magnetic metallic electrodes, where
$a_{{\bf k},\sigma}^{\alpha\dagger}$ and $a_{{\bf k},\sigma}^{\alpha}$ are creation and annihilation operators for an electron at electrode
$\alpha$ with energy $\epsilon_{{\bf k},\sigma}^{\alpha}$, momentum ${\bf k}$ and spin $\sigma$. The Hamiltonian for the tunneling between
the electrodes $\alpha$ and the molecule is
\begin{eqnarray}
{\cal H}_{\rm T} &=& \sum_{\alpha=L,R}\sum_{{\bf k},\sigma}
(t_{\alpha}^{\star}c_{\sigma}^{\dagger}a_{{\bf k},\sigma}^{\alpha}
+ t_{\alpha}a_{{\bf k},\sigma}^{\alpha\dagger}c_{\sigma}),
\end{eqnarray}
where $t_{\alpha}^{\star}$ and $t_{\alpha}$ are tunneling parameters. We assume symmetric tunneling such that $t_{L}=t_{R}$. Since we
consider the case that there is no bonding between the molecule and the electrodes, we assume weak coupling of the molecule to the
electrodes. Thus, sequential tunneling is dominant. The tunneling parameters may, in general, depend on charge state, magnetic level,
magnetic field, or gate voltage. However, we assume that the tunneling parameters are constant, i.e. do not depend on any of them.
It is not the scope of this work to include such dependence or to estimate the values of tunneling parameters.

Considering the weak coupling between the electrodes and the molecule, ${\cal H}_{\rm T}$ is a small perturbation to ${\cal H}_{\rm el}$ and
${\cal H}_{\rm mol}$. A total wave function, $| \Psi \rangle$, of ${\cal H}={\cal H}_{\rm mol}+{\cal H}_{\rm el}+{\cal H}_{\rm T}$
can be expressed as a direct product of a wave function of electrode $\alpha$ and molecular eigenstate $|q \rangle$.
As a basis set for ${\cal H}_{\rm mol}$, we use $\{ |S=11/2, M \rangle, M=-11/2, -9/2, ..., 9/2, 11/2 \}$, where $M$ are eigenvalues of the
$z$ component of spin operator $S_z$. The molecular levels in the $N=0$ state can be written in terms of the basis set and the spinors
($|\uparrow \rangle$,$|\downarrow \rangle$) by using Clebsch-Gordon coefficients. The molecular eigenstates for state $N=-1$ ($N=0$) can
be written as $\sum_j u_j |M_j \rangle$ ($\sum_k v_k |m_k \rangle$), where $u_j$ ($v_k$) are coefficients and $M_j=-11/2, ..., 11/2$
($m_k=-6, ..., 6$).

In the sequential tunneling limit, transition rates $R_{i \rightarrow f}$ from initial state $| \Psi_i \rangle$ to
final state $| \Psi_f \rangle$, are shown as $2\pi/\hbar | \langle \Psi_f | {\cal H}_{\rm t} | \Psi_i \rangle|^2 \delta(E_f - E_i)$,
to the lowest order in ${\cal H}_{\rm T}$, where $E_f$ and $E_i$ are the final and initial energies, respectively.
The transition rates are integrated over ${\mathbf k}$ and thermal distributions of electrons in the electrodes are described by
the Fermi-Dirac distribution function $f(E)$.
For electron tunneling from electrode $\alpha$ to the molecule, the molecule undergoes a transition from level
$|q \rangle$ in the $N=-1$ state to level $|r \rangle$ in the $N=0$ state, where $|q \rangle = \sum_j u_j |M_j \rangle$ and
$|r \rangle = \sum_k v_k |m_k \rangle$. The corresponding transition rates $\gamma_{\alpha}^{q \rightarrow r}$, are given by
\cite{MISI07,MISI10,MISI09,KOCH05}
\begin{eqnarray}
\label{eq:tran01}
\gamma_{\alpha}^{q \rightarrow r} &=& \sum_{\sigma} W^{\sigma, \alpha}_{q \rightarrow r} f( \epsilon_r - \epsilon_q - \mu_{\alpha} ),
\label{eq:tran1}
\end{eqnarray}
where $W^{\sigma, \alpha}_{q \rightarrow r}$ is
$\frac{2 \pi}{\hbar} {\cal D}^{\alpha}_{\sigma} |t_{\alpha}|^2 | \langle r | c^{\dag}_{\sigma} | q \rangle |^2$, and
${\cal D}^{\alpha}_{\sigma}$ is density of states of electrode ${\alpha}$ near the Fermi level ($E_F$). We assume that
${\cal D}^{\alpha}_{\sigma}$ is constant and that it is independent of $\alpha$ and $\sigma$. Here $\mu_{\alpha}$ is a chemical
potential of electrode $\alpha$. Considering a symmetric bias application, we set $\mu_L = eV/2$ and $\mu_R = -eV/2$, where
$V$ is a bias voltage. The matrix elements $\langle r | c^{\dag}_{\sigma} | q \rangle$ determine selection rules such as
$|\Delta S|=1/2$, $|M_j - m_k|=1/2$, and $\Delta N=\pm 1$. For tunneling from the molecule to electrode $\alpha$, the
molecule make a transition from $|r \rangle$ to $|q \rangle$. The transition rates are now given by
\begin{eqnarray}
\label{eq:W2}
\gamma_{\alpha}^{r \rightarrow q} &=& \sum_{\sigma} W^{\sigma, \alpha}_{r \rightarrow q}
[1 - f( \epsilon_r - \epsilon_q - \mu_{\alpha} )],
\label{eq:tran2}
\end{eqnarray}
where $[1 - f( \epsilon_r - \epsilon_q - \mu_{\alpha} )]$ is included because the final state
must be unoccupied for an electron to tunnel back to electrode $\alpha$.

\begin{figure}
    \includegraphics[width=0.22\textwidth]{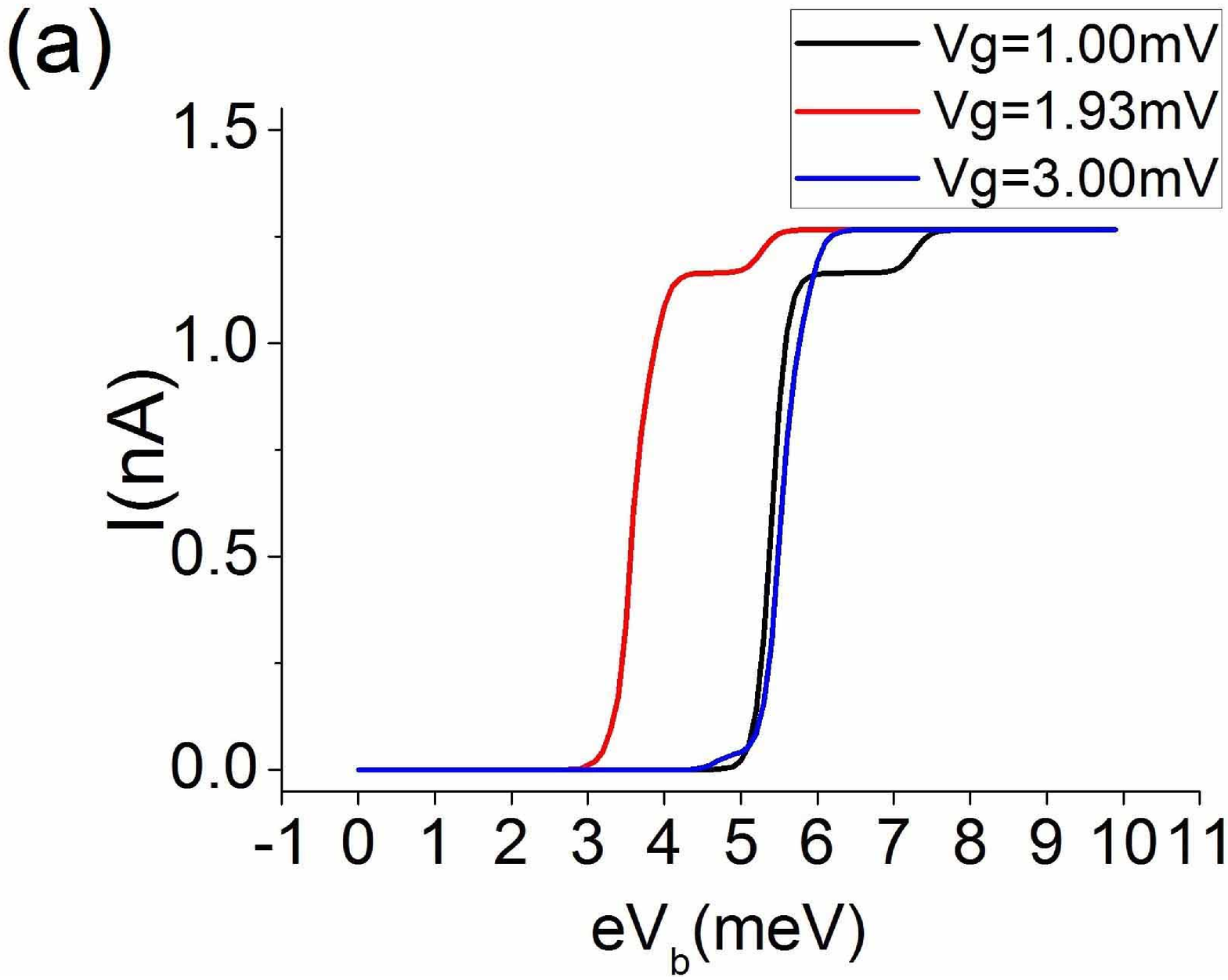}
    \includegraphics[width=0.25\textwidth]{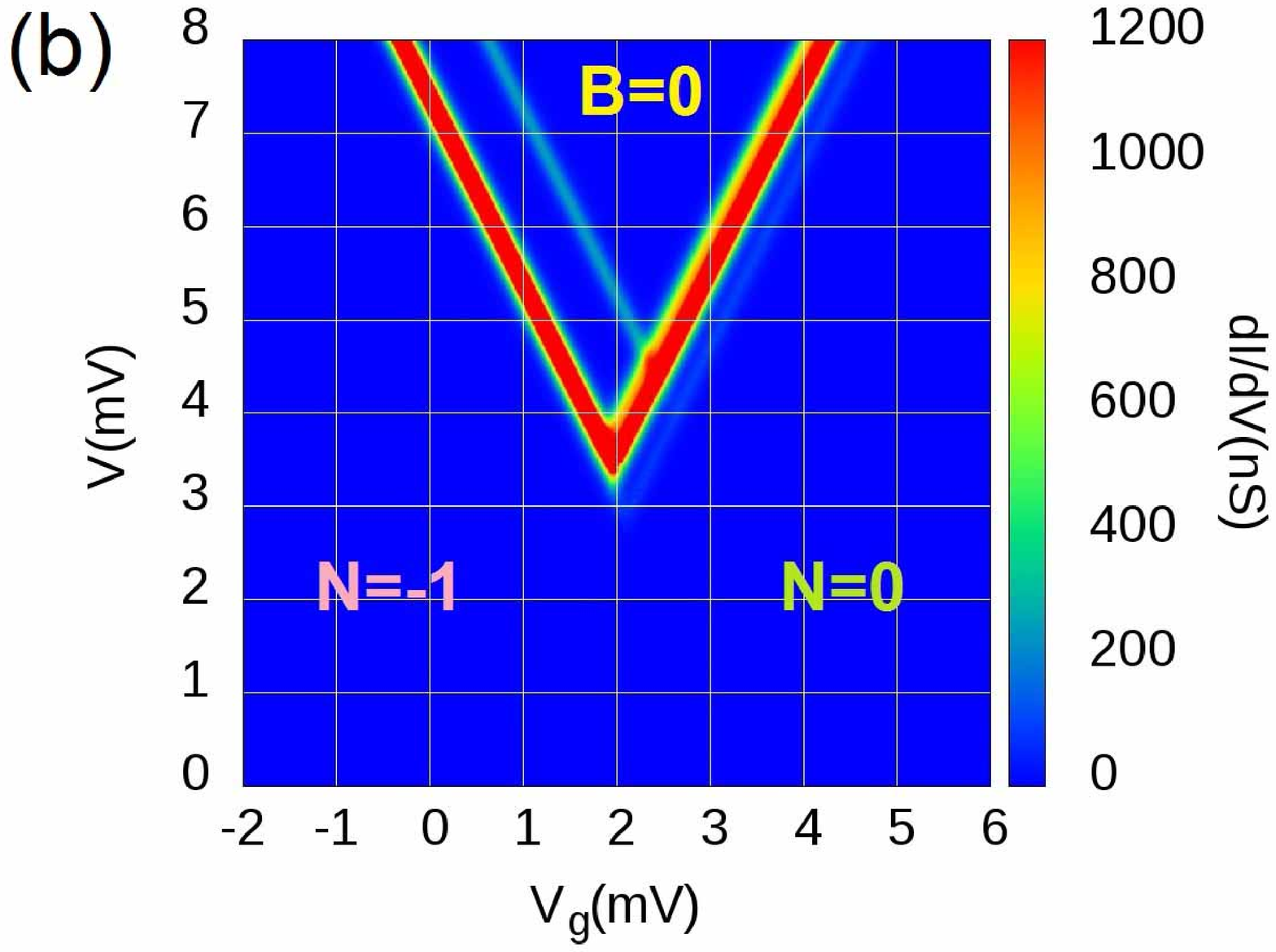}
    \includegraphics[width=0.235\textwidth]{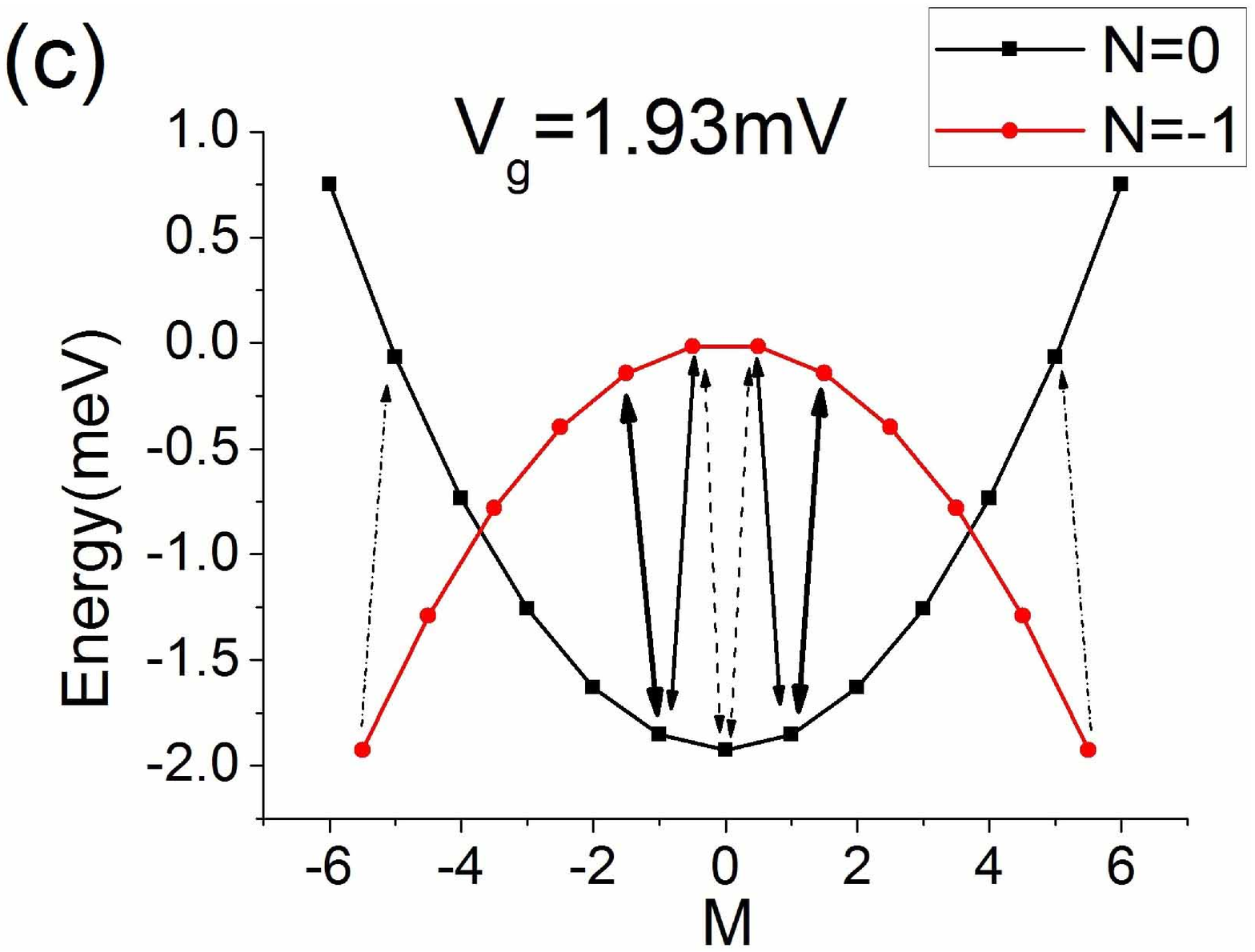}
    \includegraphics[width=0.235\textwidth]{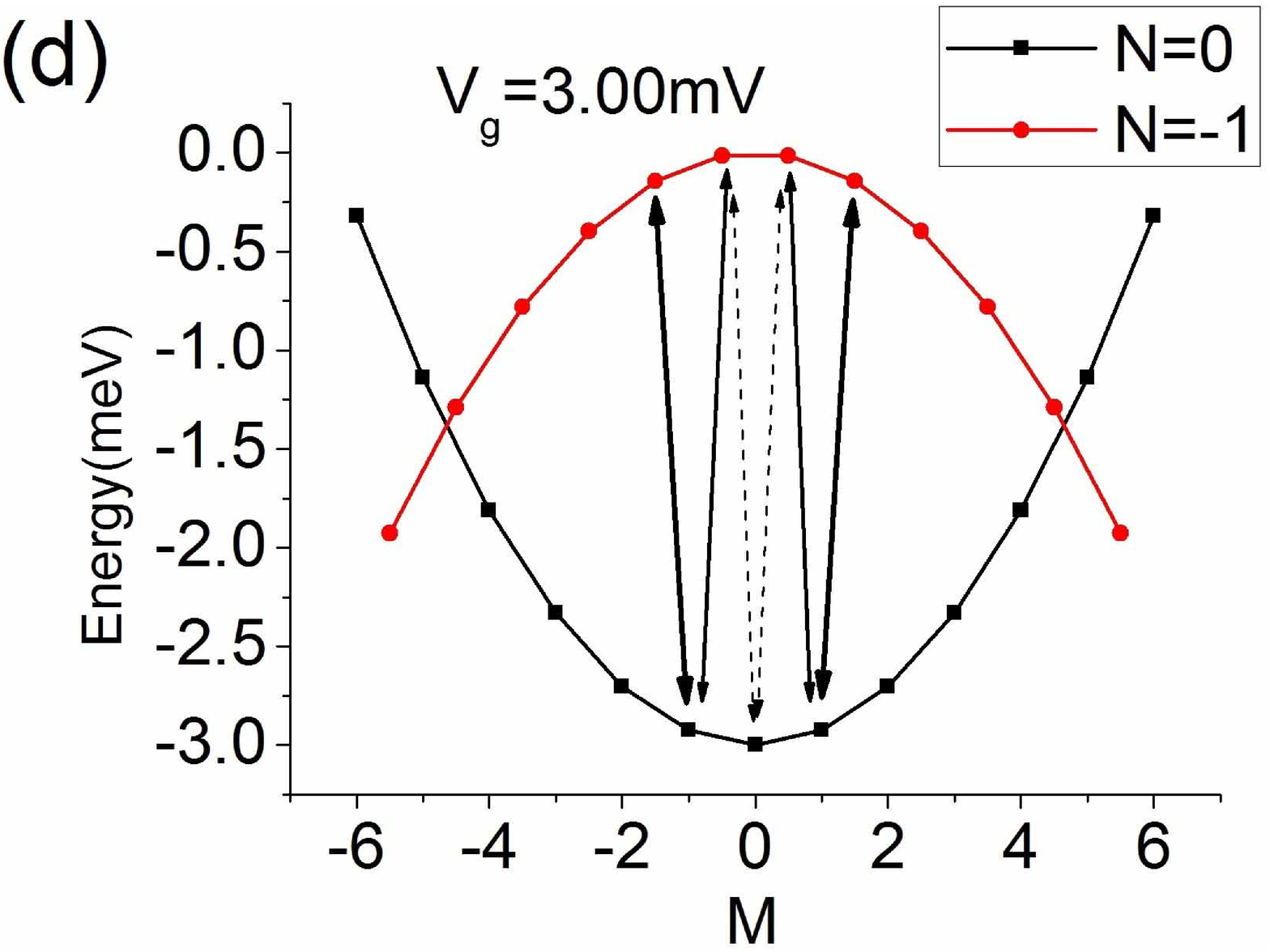}
\caption{(Color online) (a) Calculated $I$-$V$ curves at $V_g=1.0$, 1.93, and 3.0~mV for $T$$=$0.05~meV/$k_B$.
(b) Calculated $dI/dV$ as a function of $V$ and $V_g$ for ${\mathbf B}=0$. The magnetic levels
in the $N$$=$0 and $N$$=$1 states for
(c) $V_{g}=1.93$ mV and (d) $V_g=3.0$~mV with ${\mathbf B}=0$. Arrows in (c) and (d) indicate transitions where
the energy differences between the involved levels are comparable to $eV/2$ for the first $dI/dV$ peak.}
\label{fig:Ebasic}
\end{figure}

\begin{figure}
    \includegraphics[width=0.35\textwidth]{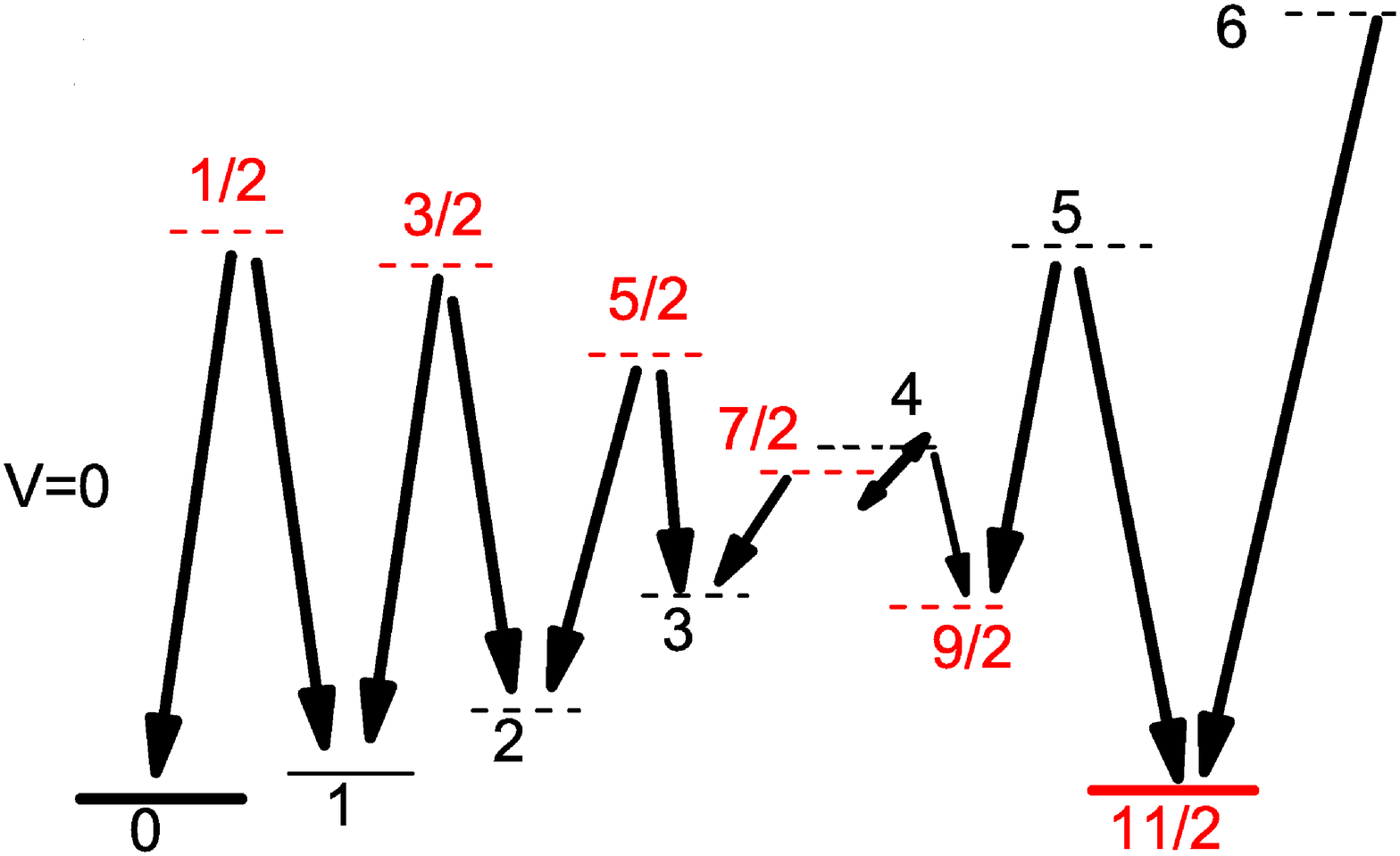}
\caption{(Color online) Schematic diagram to elucidate the mechanism for the spin blockade effect at
$B=0$ for $V=0$~mV and $V_g=1.93$~mV. Only the $m$ and $M$ levels with $m, M > 0$ are shown since
the $m, M < 0$ levels are degenerate with the $m, M > 0$ levels. The thicknesses of the levels represent the values of the
level occupation probabilities: thicker lines for higher occupation probabilities. The dashed lines indicate that their occupation
probabilities are zero. Thicker arrows are for higher transition rates. Only non-zero transition rates are shown. None of the
indicated transitions contribute to the current at zero bias. See the text for details.}
\label{fig:scheme}
\end{figure}

Probabilities $P_q$ of molecular states $|q \rangle$ being occupied must satisfy the following master
equation \cite{MISI07,MCCA15,MISI10,MISI09,KOCH05}
\begin{eqnarray}
\frac{dP_q}{dt} &=& - P_q \sum_{\alpha=L,R} \sum_{r} \gamma_{\alpha}^{q \rightarrow r}
+ \sum_{\alpha=L,R} \sum_{r} \gamma_{\alpha}^{r \rightarrow q} P_r,
\label{eq:master}
\end{eqnarray}
where the summation over $r$ runs for the orbital and magnetic degrees of freedom. The first (second) term covers all allowed transitions
from (to) $| q \rangle$. To find steady-state probabilities, we solve $dP_q/dt = 0$ for $P_q$, by applying the bi-conjugate gradient
stabilized algorithm \cite{SLEI93} with the zero-bias Boltzmann distribution as initial occupation probabilities. Then the current
from electrode $\alpha$ to the molecule is computed from
\begin{eqnarray}
I_{\alpha=L,R} &=& e \sum_{q,r} \gamma_{\alpha}^{|N=-1,q \rangle \rightarrow |N=0,r \rangle} P_q \\ \nonumber
 & & - e \sum_{q,r} \gamma_{\alpha}^{|N=0,r \rangle \rightarrow |N=-1,q \rangle} P_r,
\end{eqnarray}
where the sums over $q$ and $r$ run for all orbital and magnetic indices. The total current is obtained from $I=(I_L-I_R)/2$.
A differential conductance ($dI/dV$) is numerically calculated from current-voltage ($I-V$) characteristics by using a small bias
interval $\Delta V \le 0.1$~mV.

\section{Results and Discussion}

\begin{figure*}
    \includegraphics[width=0.34\textwidth]{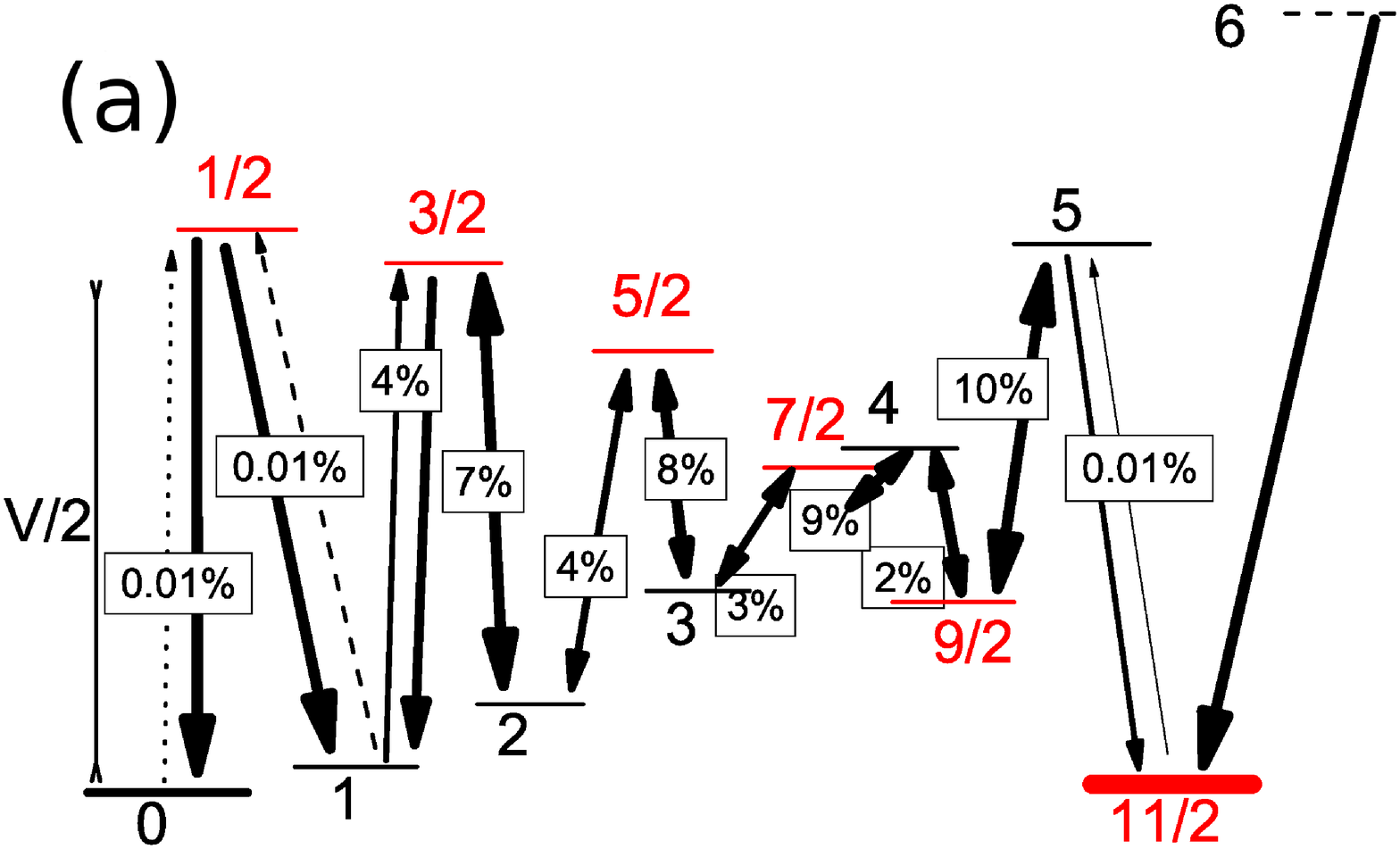}
    \hspace{0.2truecm}
    \includegraphics[width=0.3\textwidth]{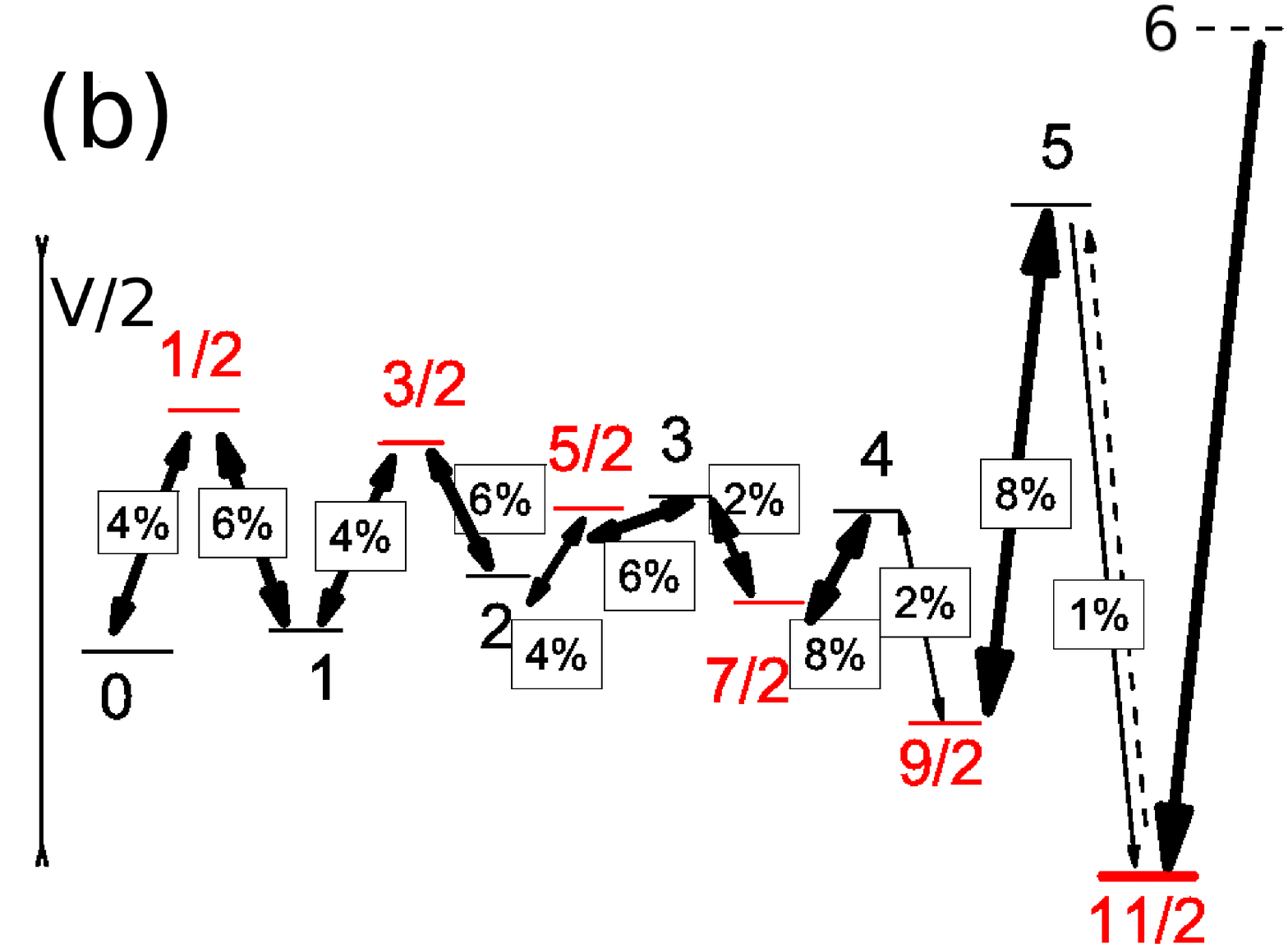}
    \hspace{0.2truecm}
    \includegraphics[width=0.3\textwidth]{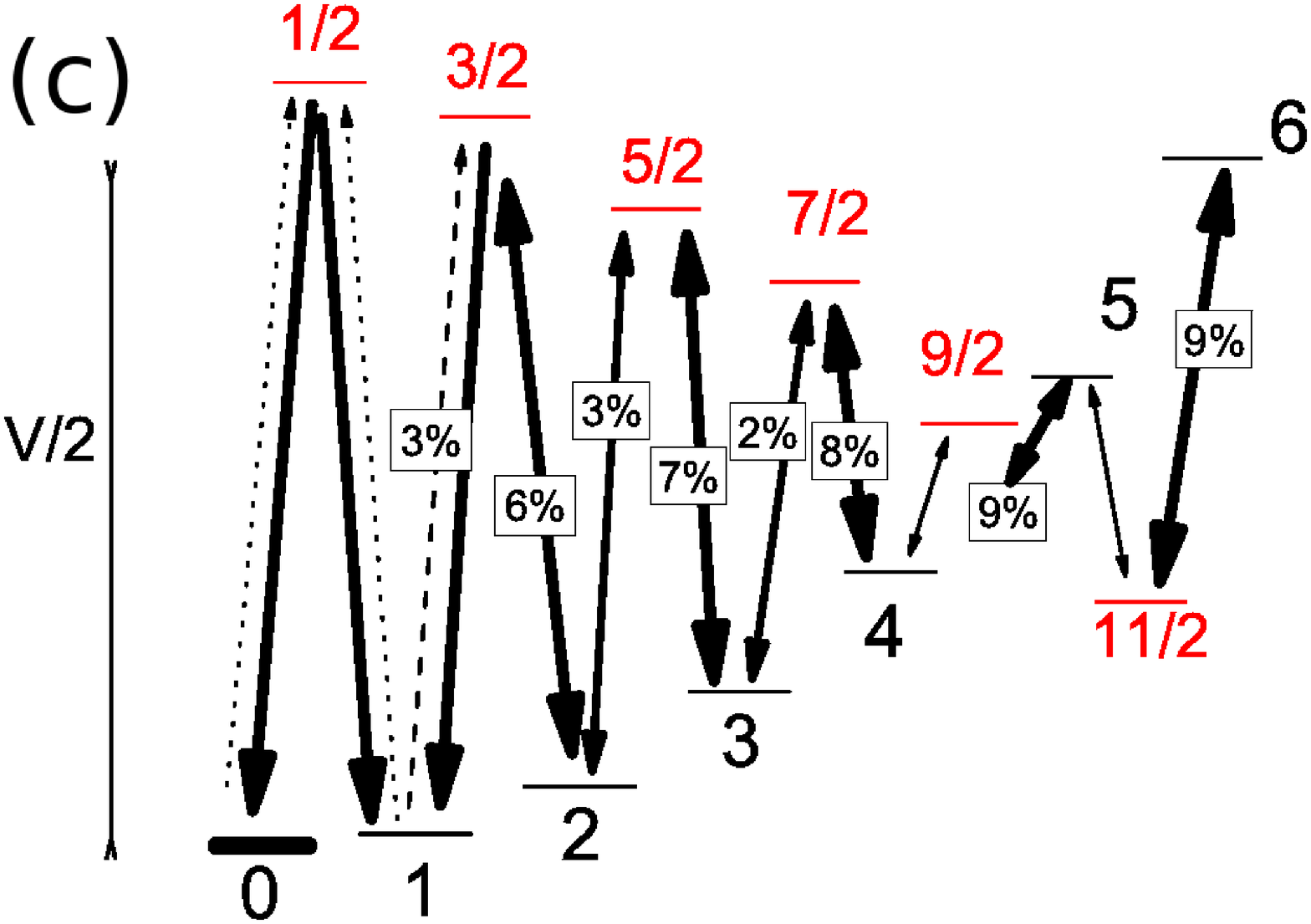}
\caption{(Color online) Schematic diagrams to elucidate the mechanism for the lifting of the spin blockade effect at
$B=0$ for (a) $V=3.4$~mV and $V_g=1.93$~mV, (b) $V=5.4$~mV and $V_g=1.0$~mV, and (c) $V=5.4$~mV and $V_g=3.0$~mV.
Only the $m$ and $M$ levels with $m, M > 0$ are shown. The thicknesses of the levels represent the values of the occupation
probabilities: thicker lines for higher occupation probabilities. The dashed horizontal lines indicate that the occupation
probabilities are zero. Thicker arrows are for higher transition rates and dashed arrows are for smallest transition rates.
Only non-zero transition rates are shown. Boxed numbers indicate percentage contributions of particular transition
pathways to the current. For transitions which do not contribute to the current, boxed numbers are absent. The bias
window is shown.}
\label{fig:B0occ}
\end{figure*}

We present calculated $dI/dV$ as a function of $V$ and $V_g$ at temperature $T=0.58$~K (or 0.05~meV/$k_B$) for ${\mathbf B}=0$
and for ${\mathbf B}$ field along the $z$ axis and the $x$ axis, separately. In all cases, the level broadening parameter,
$\Gamma=2\pi {\cal D} |t|^2$, is set to be 0.01~meV such that $\Gamma \ll k_B T$, where ${\cal D}$ is density of states of the
electrodes and $k_B$ is the Boltzmann constant. Only sequential tunneling is considered since it is dominant. Our work
qualitatively differs from the previous work on the molecule \cite{KEXU09} because the magnetic levels were not
considered in the latter.

\subsection{Spin blockade without ${\mathbf B}$ field}

Figure~\ref{fig:Ebasic}(a) shows our calculated $I$-$V$ curves at three gate voltages for ${\mathbf B}=0$. At $V_g=1.93$~mV
current flows through the molecule only for $V \geq 3.4$~mV, and it increases further near 5.4~mV with a small step.
At $V_g=1.0$ and 3.0~mV current flows for $V \geq 5.3$~mV. In the three cases current is blocked until $V$ reaches a certain value.
The same feature is found at any other gate voltages, as shown in Fig.~\ref{fig:Ebasic}(b). This current suppression at low bias
is the essence of our work, i.e. the magnetic-anisotropy induced spin blockade effect.

With $V_g=1.93$~mV the molecule is at the charge degeneracy point (CDP), where the lowest levels in the two charge states,
$|N=-1;S=11/2,M=\pm11/2 \rangle$ and $|N=0;S=6,m=0 \rangle$, are degenerate as shown in Fig.~\ref{fig:Ebasic}(c).
At $V=0$, in the $N=-1$ state only the $M=\pm 11/2$ levels are occupied with probabilities of 29 and 29\%, while in the $N=0$ state the
$m=0$ and $m=\pm 1$ levels have occupation probabilities of 29, 6.5, and 6.5\%, respectively. The rest of the levels are not occupied
because there are zero rates of transitions from the occupied levels to higher-energy levels, although the transitions are allowed
by the selection rules. (Only rates of the reverse transitions are quite high.) See Fig.~\ref{fig:scheme}].
The former transition rates vanish since there are no occupied electrons at the corresponding energy levels in the electrodes according
to the Fermi-Dirac distribution $f(E)$. Furthermore, there are no allowed transitions among the occupied levels. Therefore, no current
flows at zero bias. Increasing the bias, some higher levels enter the bias window.
Due to the symmetric bias application, only the levels whose energy differences are less than about $eV/2$ can contribute to the current.
For example, at $V=2.3$~mV, energy $eV/2$ is close to the energy difference between $M=\pm 5/2$ and $m=\pm 2$
[Figs.~\ref{fig:Ebasic}(c) and \ref{fig:scheme}]. Thus, at $V=2.3$~mV, transitions such as ($|M=\pm7/2 \rangle$, $|m=\pm3 \rangle$),
($|m=\pm4 \rangle$, $|M=\pm7/2 \rangle$), ($|m=\pm4 \rangle$, $|M=\pm9/2 \rangle$), and ($|M=\pm5/2 \rangle$, $|m=\pm2 \rangle$), are
allowed by the selection rules and possible in terms of energy differences.
However, higher energy levels $m=\pm 2, \pm 3, \pm 4$ and $M=\pm 5/2, \pm 7/2, \pm 9/2$ are not occupied since there are no transition
pathways to occupy them from the zero-bias occupied levels at this bias voltage. Therefore, these higher levels cannot contribute to
the current. Only when there exist significant transition rates {\it from} at least one of the initial (zero-bias) five occupied
levels $M=\pm 11/2$ and $m=0, \pm 1$ to higher levels, the higher-energy levels are occupied. Then the occupied levels can contribute
to the current and they induce other transitions. Figure~\ref{fig:B0occ}(a) shows occupation probabilities, transition rates, and
percentage contributions to the current at $V=3.4$~mV for the positive $m$ and $M$ levels including $m=0$. Note that the negative $m$ and
$M$ levels are degenerate
with the positive $m$ and $M$ levels. At this bias, the energy differences between $m=\pm1$ and $M=\pm 1/2$ or $M=\pm 3/2$ are comparable
to $eV/2$. All $m$ levels except for $m=\pm6$ and all $M$ levels are within the bias window. Significant transition rates from $m=\pm1$
to $M=\pm3/2$ make the $M=\pm3/2$ level occupied. This triggers subsequent transitions such as
$|M=\pm3/2 \rangle \leftrightarrow |m=\pm2 \rangle \leftrightarrow |M=\pm5/2 \rangle \leftrightarrow |m=\pm3 \rangle
\leftrightarrow |M=\pm7/2 \rangle \leftrightarrow |m=\pm4 \rangle \leftrightarrow |M=\pm9/2 \rangle \leftrightarrow |m=\pm5 \rangle$,
as shown in Fig.~\ref{fig:B0occ}(a). There exist also small transition rates from $M=\pm11/2$ to $m=\pm5$. All of these transitions
enable all $m$ levels except for $m=\pm 6$ and all $M$ levels to be occupied to some extent. Therefore, these transitions can contribute
to the large step near $V=3.4$~mV in the $I$-$V$ curve [Fig.~\ref{fig:Ebasic}(a)].
When the bias increases slightly more like $V=3.68$~mV, transitions from $m=0$ to $M=\pm 1/2$ increase, which makes transitions
such as ($|M=\pm1/2 \rangle$, $|m=1 \rangle$) and ($|m=\pm5 \rangle$, $|M=\pm11/2 \rangle$) additionally contribute to the current.
Now the small second step near $V=5.4$~mV in the $I-V$ plot can be explained by allowed transitions from $M=\pm11/2$ to $m=\pm6$,
where the energy difference between the levels is 2.68~meV.

\begin{figure*}
    \includegraphics[width=0.325\textwidth]{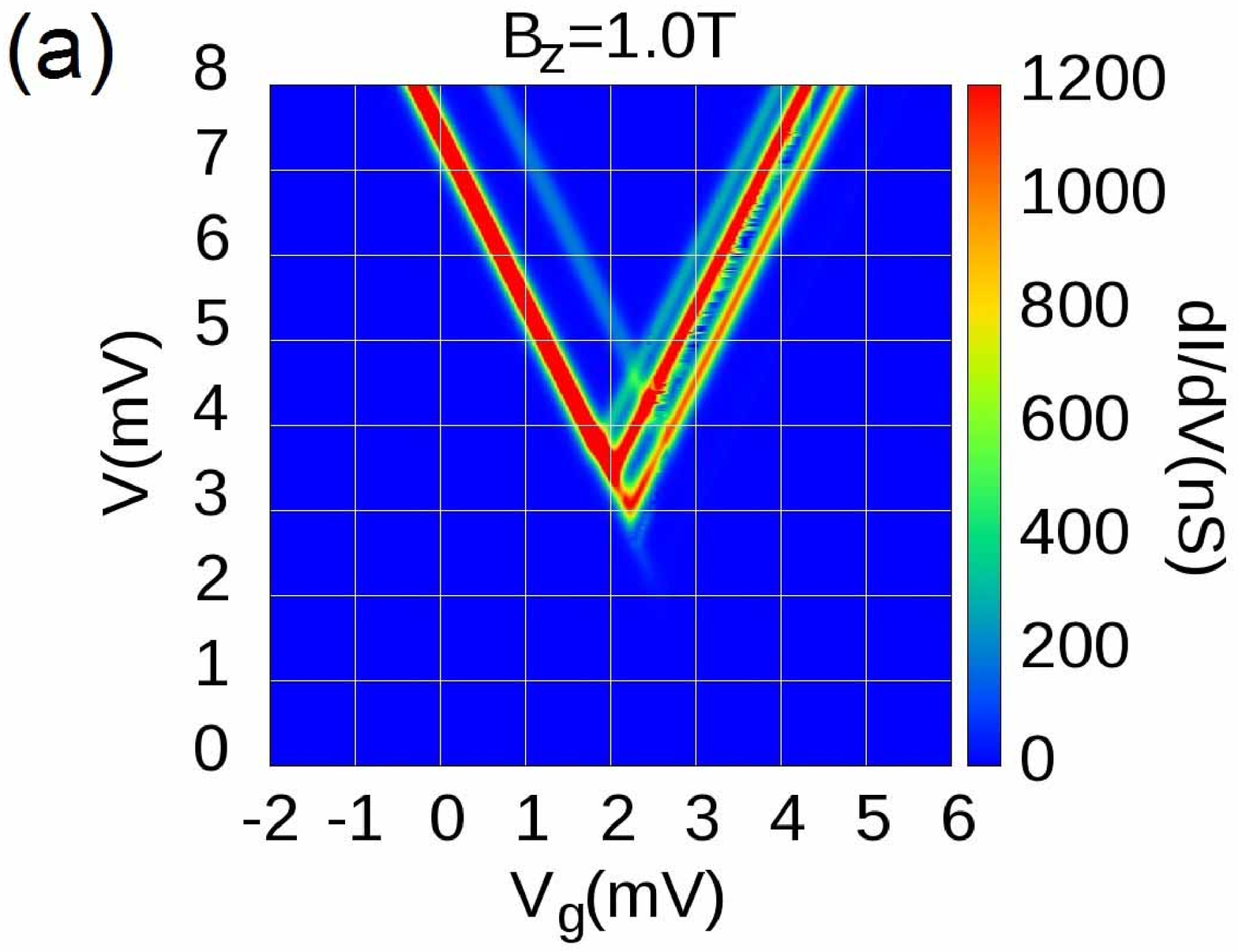}
    \includegraphics[width=0.325\textwidth]{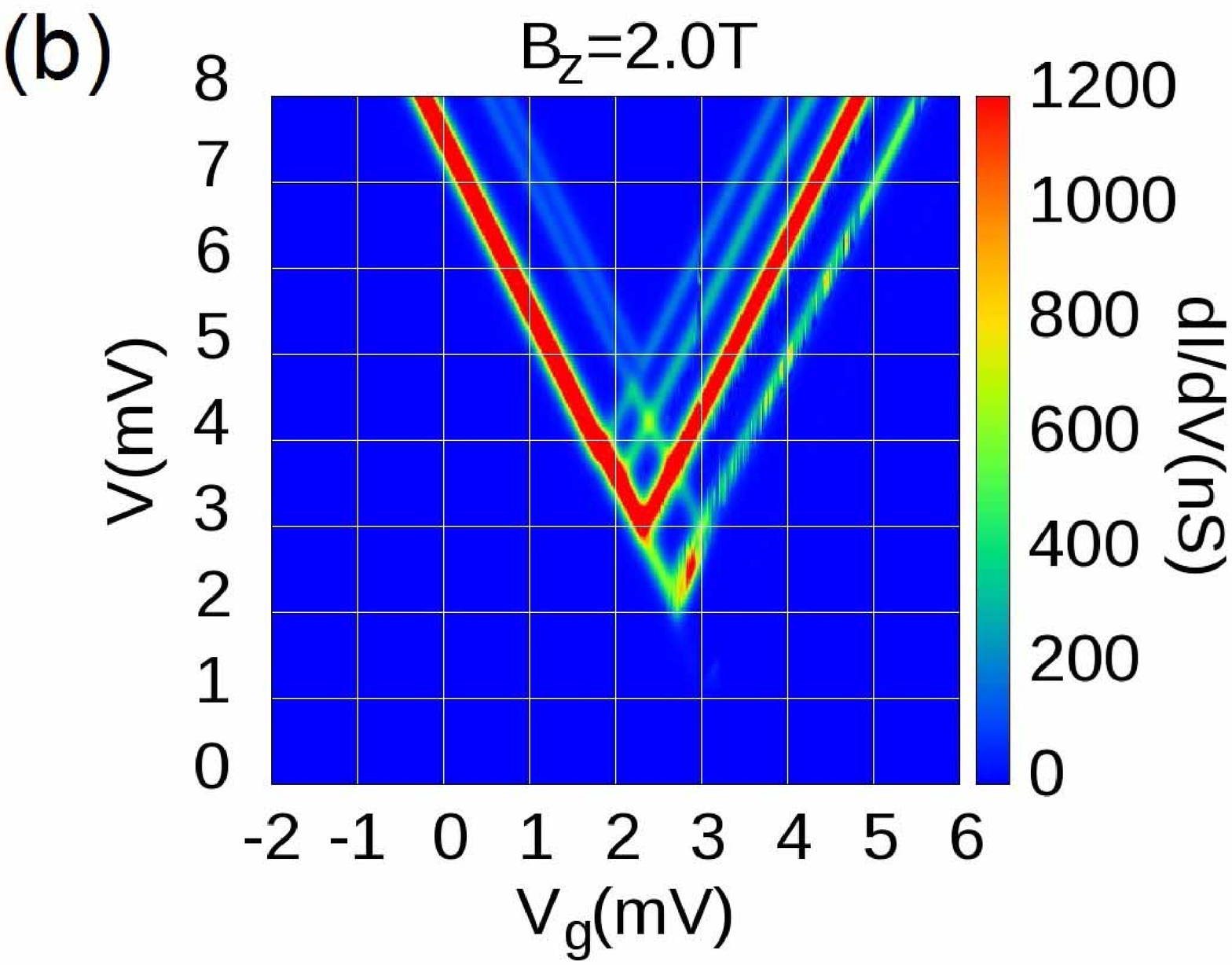}
    \includegraphics[width=0.325\textwidth]{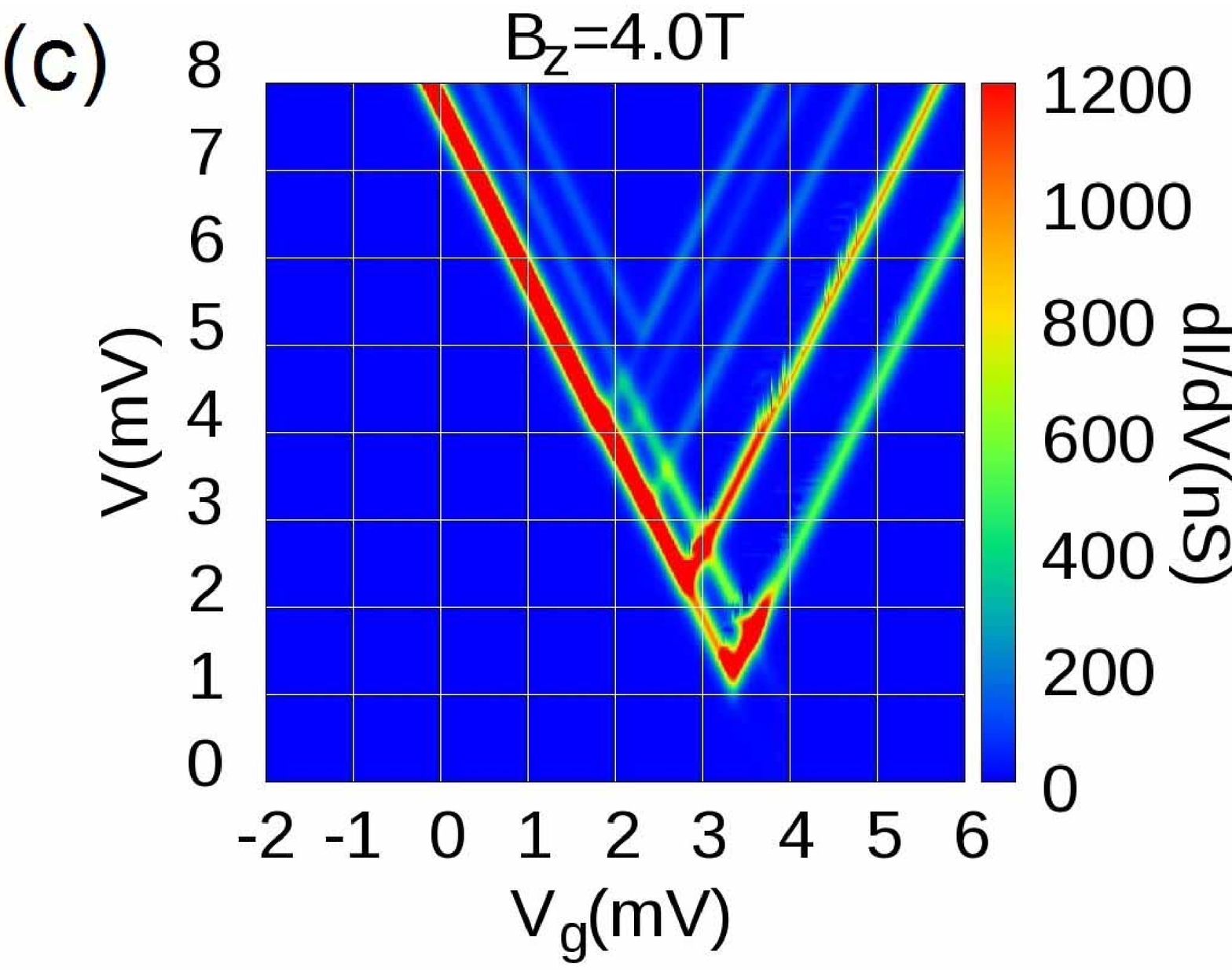}
    \includegraphics[width=0.325\textwidth]{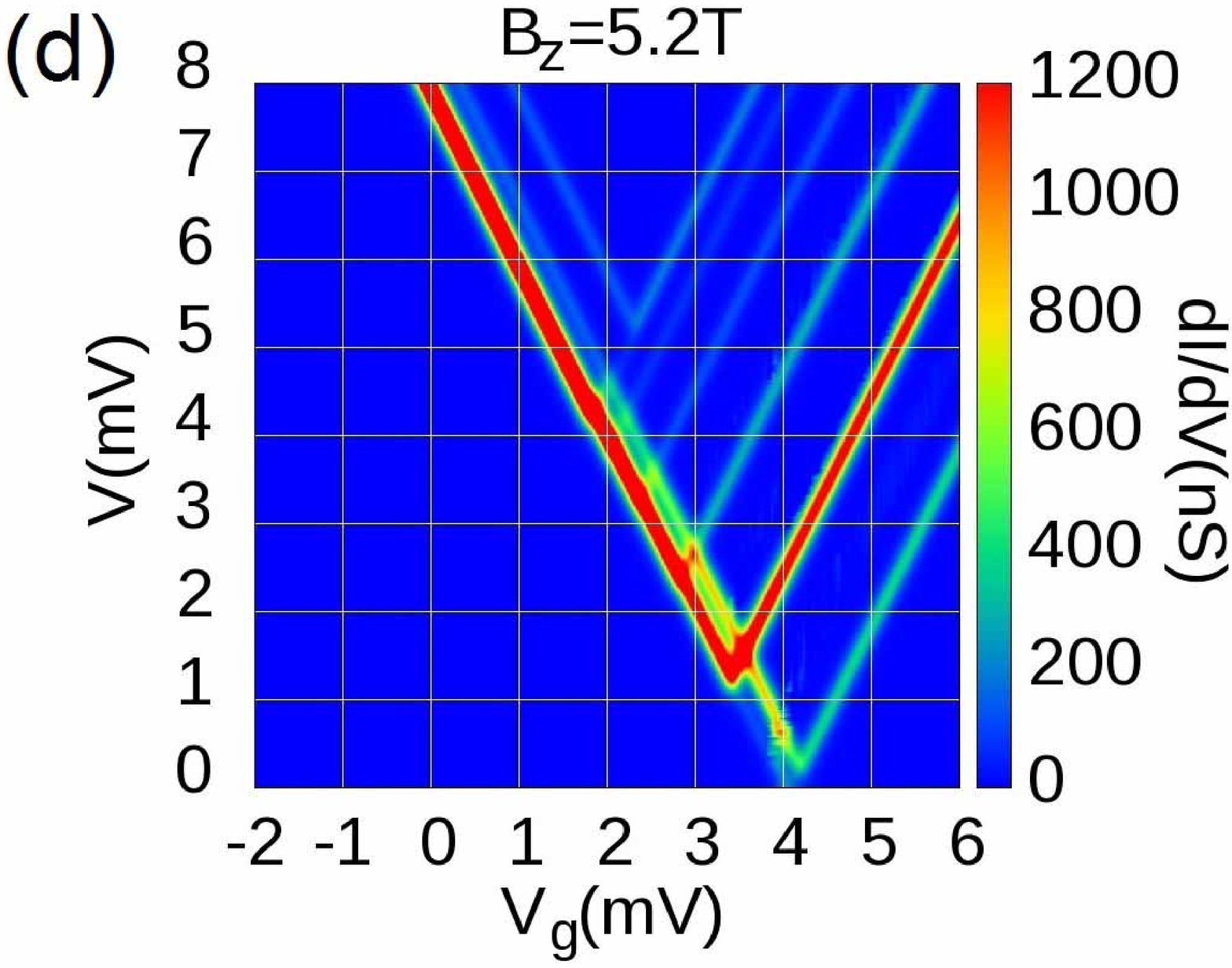}
    \includegraphics[width=0.3\textwidth]{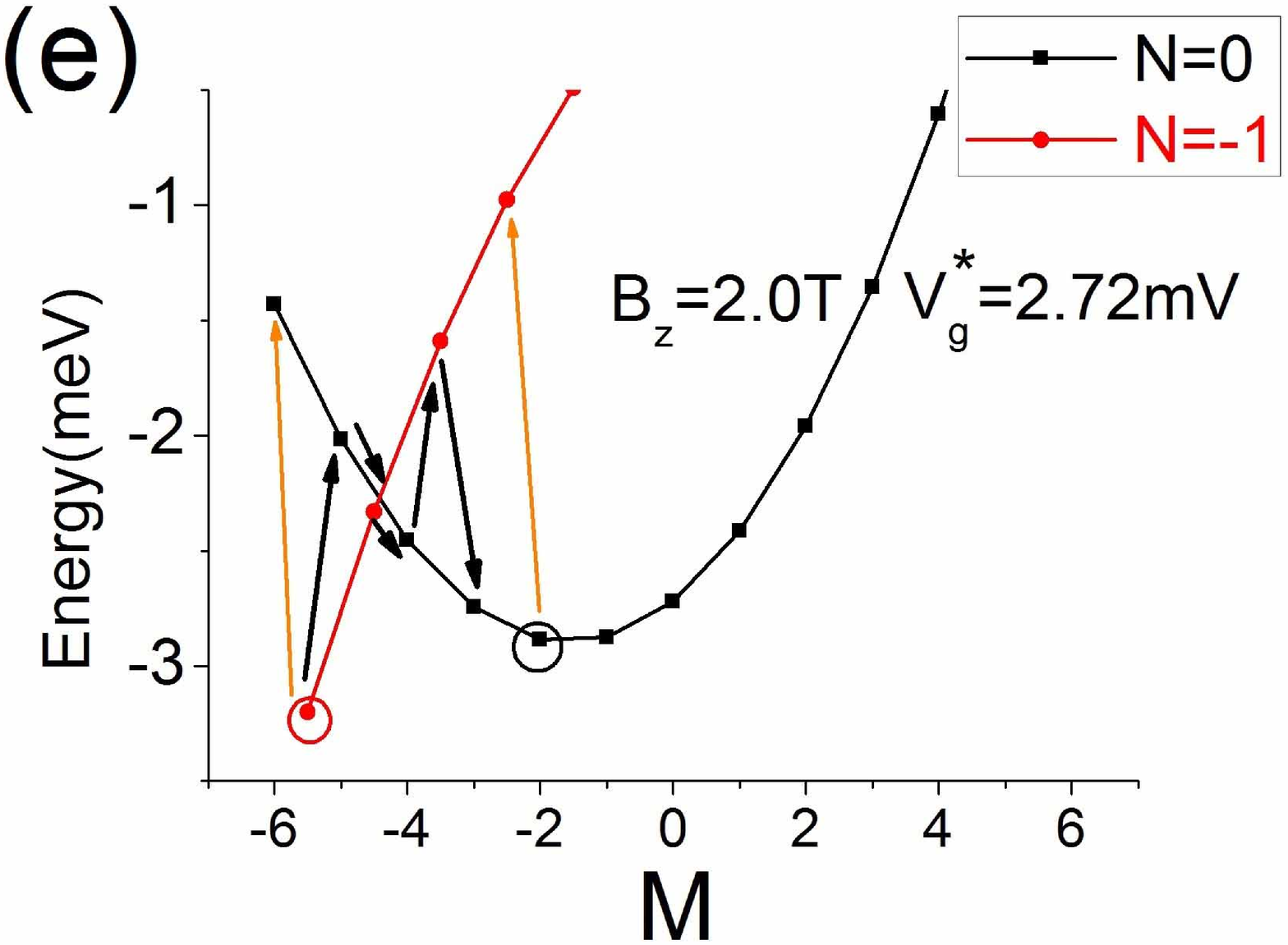}
    \vspace{0.2truecm}
    \includegraphics[width=0.3\textwidth]{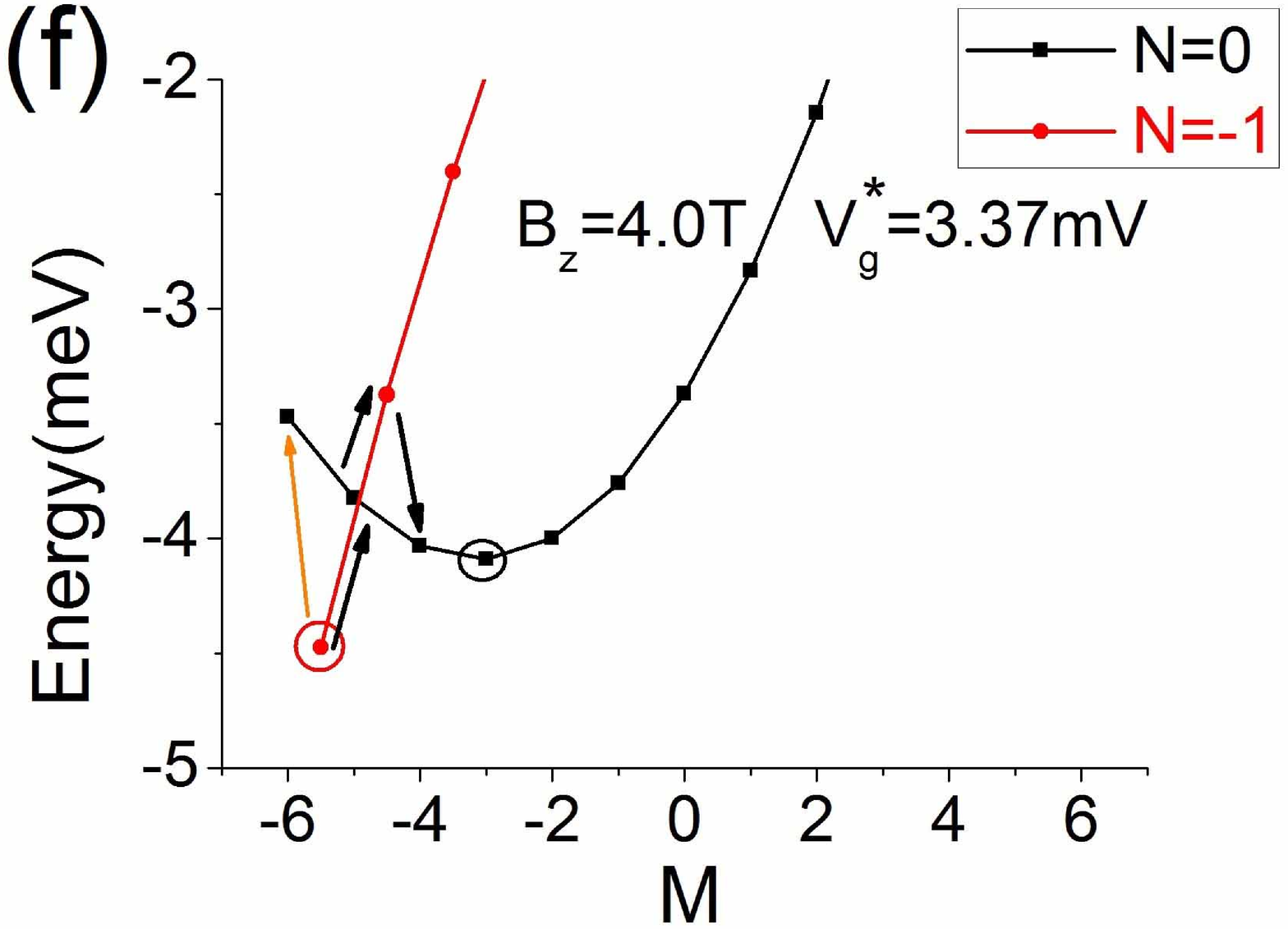}
\caption{(Color online) (a)-(d) Calculated $dI/dV$ values as a function of $V$ and $V_g$ at $T=0.05$~meV/$k_B$ for several
$B_z$ values. Magnetic level energies in the two charge states (e) for $B_z=2.0$~T at $V_g^{\star}=2.72$~mV
and for (f) $B_z=4.0$~T at $V_g^{\star}=3.37$~mV. Not all the levels are shown in (e) and (f). The thick black (thin red) arrows in (e)
and (f) indicate allowed transitions for the first (second) $dI/dV$ peak and the circled levels are the lowest levels for the $N=-1$
and $N=0$ states. See the text for details.}
\label{fig:Bz}
\end{figure*}

\begin{figure}
    \includegraphics[width=0.19\textwidth]{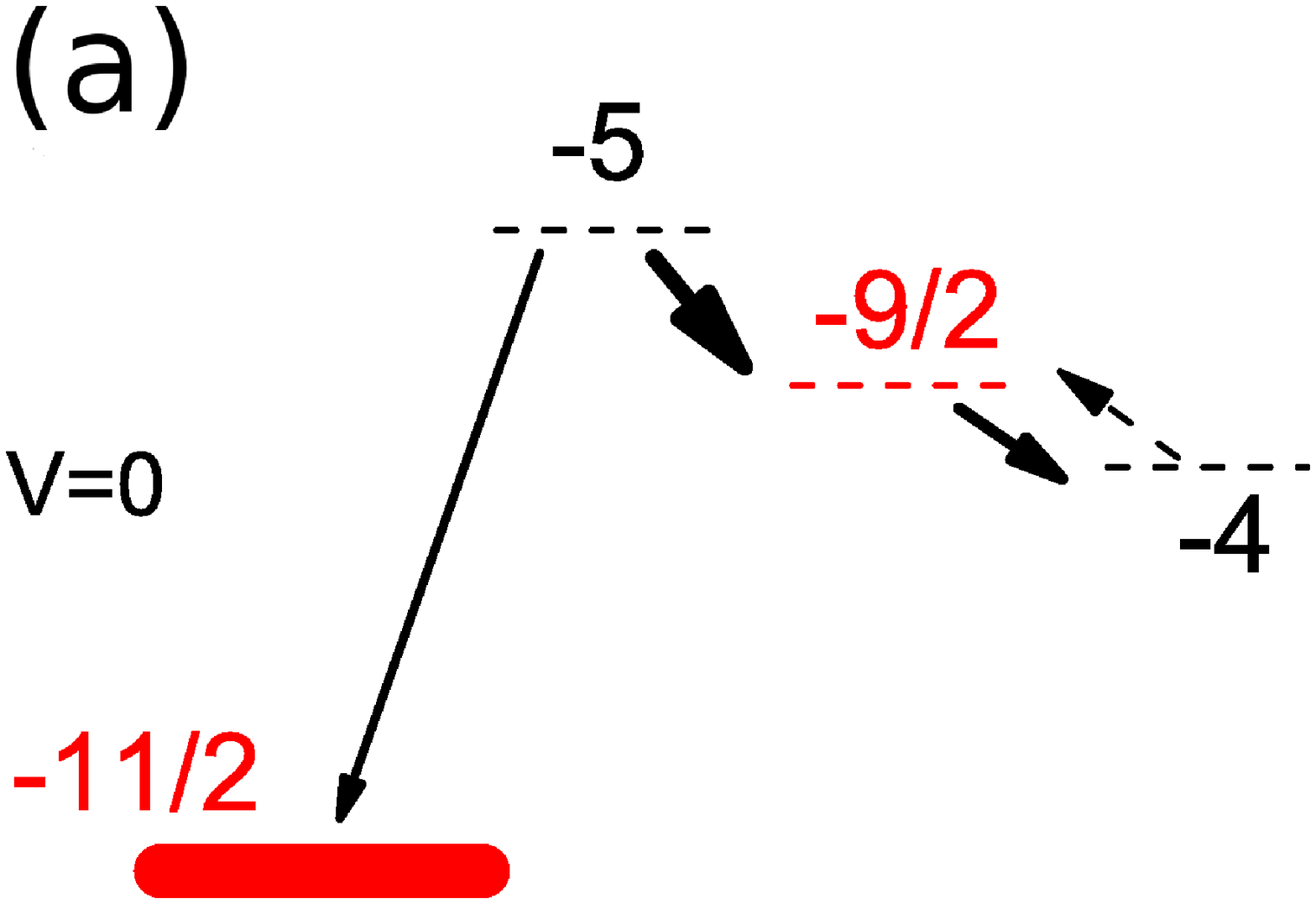}
    \hspace{0.2truecm}
    \includegraphics[width=0.25\textwidth]{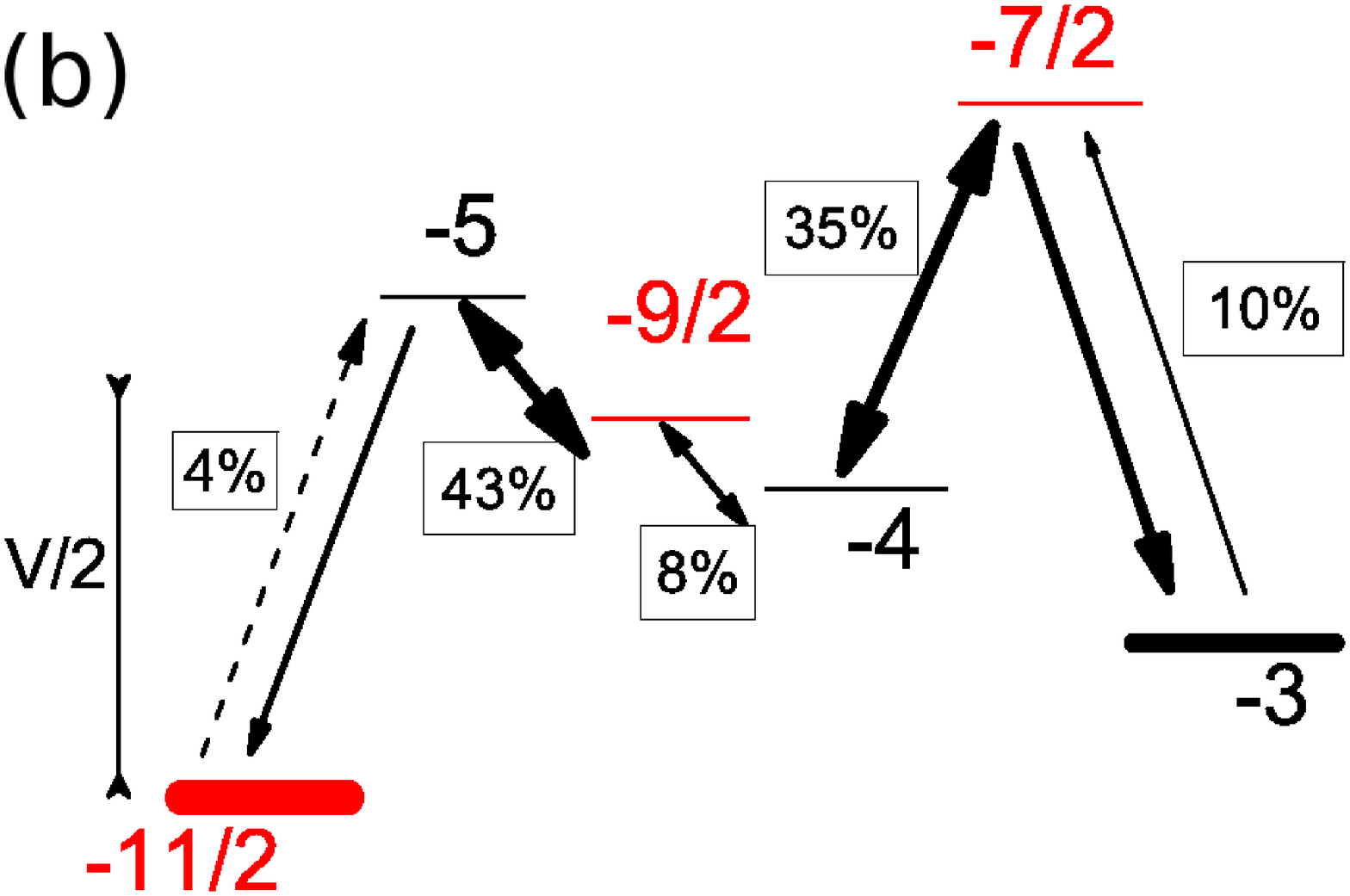}
     \includegraphics[width=0.19\textwidth]{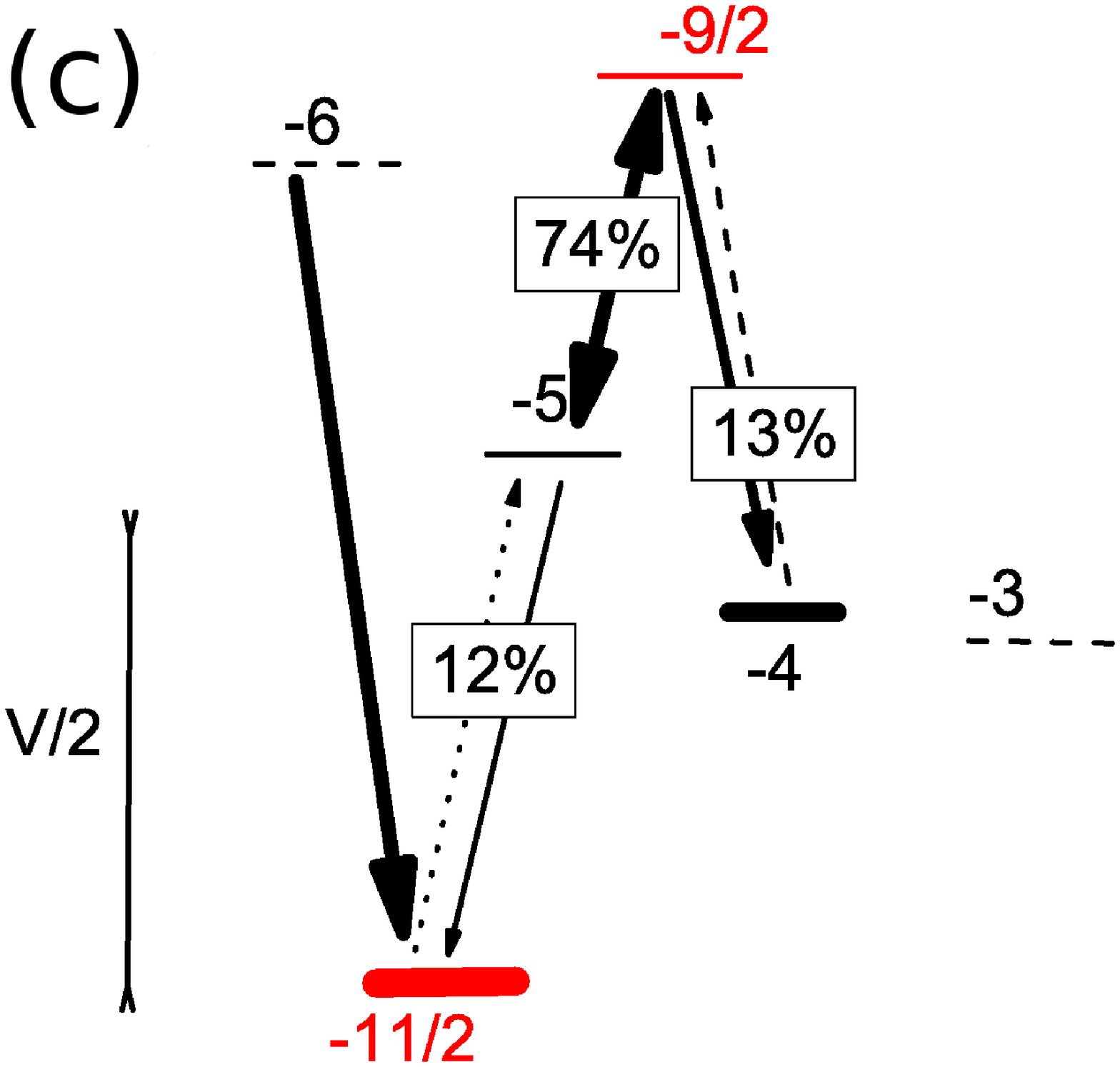}
    \includegraphics[width=0.27\textwidth]{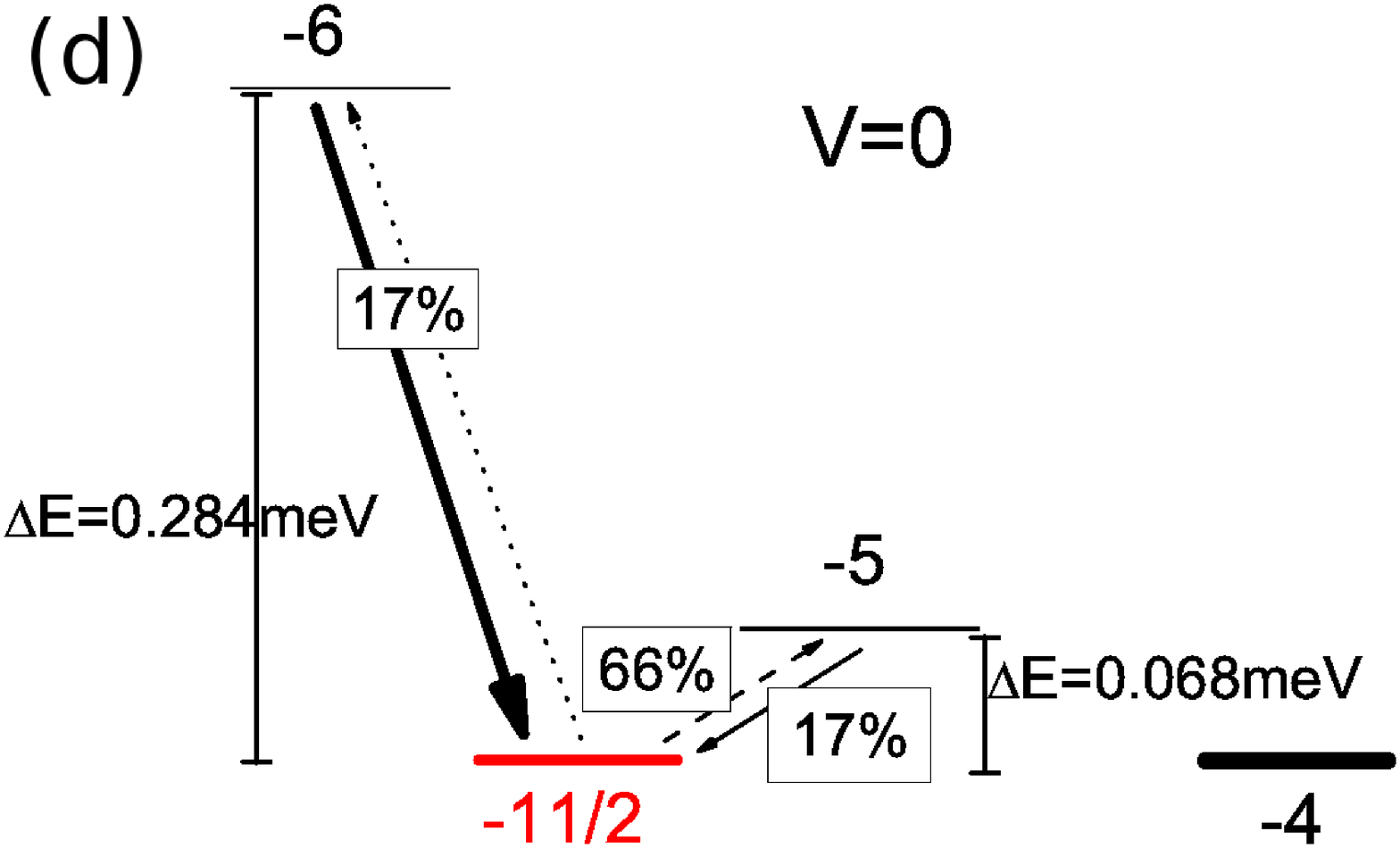}
\caption{(Color online) Schematic diagrams to elucidate the mechanisms for the spin blockade effect and for the lifting of the
effect at $B_z=2.0$~T and $V_g^{\star}=2.72$~mV for (a) $V=0$ and (b) $V=2.0$~mV, and (c) at $B_z=4.0$~T for $V_g=3.37$~mV and $V=1.0$~mV
and (d) at $B_z=5.2$~T for $V_g=4.02$~mV and $V=0$. The energy levels in (a) and (b) are zoom-in of Fig.~\ref{fig:Bz}(e),
while the energy levels in (c) are zoom-in of Fig.~\ref{fig:Bz}(f). Only relevant levels in the $N=-1$ (red) and $N=0$ (black)
states are shown. The levels not shown have extremely small or zero occupation probabilities or not relevant to our discussion.
The thicknesses of the levels represent the values of the occupation probabilities: thicker lines for higher occupation
probabilities. The dashed horizontal lines indicate that the occupation probabilities are zero. Thicker arrows are for higher
transition rates and dashed arrows are for smallest transition rates. Only non-zero transition rates are shown. Boxed numbers
indicate percentage contributions of particular transition pathways to the current. The bias window is shown. The vertical energy
scales in (a) and (b) are the same, but they differ from those in (c) and (d).}
\label{fig:occ}
\end{figure}

At $V_g=1.0$~mV the molecule is in the cationic state (at $|S=11/2,M=\pm11/2 \rangle$) for zero bias. The energy levels at this gate
voltage are shown in Fig.~\ref{fig:B0occ}(b). In this case, high current is expected only above $V=5.3$~mV because there are no transition
pathways from the zero-bias occupied levels ($M=\pm11/2$) to allowed levels in state $N=0$ at bias below $V=5.3$~mV.
At $V=5.4$~mV, there are small rates of transitions from $M=\pm 11/2$ to $m=\pm5$, and all $M$ and $m$ levels
except for $m=\pm6$ are within the bias window. See Figs.~\ref{fig:Ebasic}(a) and ~\ref{fig:B0occ}(b) Thus, these transitions induce
a series of transitions such as
$|m=\pm5 \rangle \leftrightarrow |M=\pm9/2 \rangle \leftrightarrow |m=\pm4 \rangle \leftrightarrow |M=\pm7/2 \rangle
\leftrightarrow |m=\pm3 \rangle \leftrightarrow |M=\pm5/2 \rangle \leftrightarrow |m=\pm2 \rangle \leftrightarrow
|M=\pm3/2 \rangle \leftrightarrow |m=\pm1 \rangle \leftrightarrow |M=\pm1/2 \rangle \leftrightarrow |m=0 \rangle$. As a result,
the levels within the bias window are all occupied with some probabilities. Therefore, these levels contribute to the high current.
The second small step in the $I$-$V$ curve is caused by the transitions from $M=\pm11/2$ to $m=\pm6$, similarly to the case
of $V_g=1.93$~mV,.

Now at $V_g=3.0$~mV the molecule is in the neutral state with probabilities of 68.8, 15.6, and 15.6\% at the $m=0$ and $m=\pm1$ levels,
respectively, at zero bias. The energy levels at this gate voltage are shown in Fig.~\ref{fig:B0occ}(c). In this case, only one giant step
appears in current near $V=5.3$~mV [Fig.~\ref{fig:Ebasic}(a) and (d)] since there are no transition pathways from the zero-bias occupied
levels ($m=0$ and $m=\pm1$) to allowed levels in state $N=-1$ at bias below $V=5.3$~mV. At $V=5.4$~mV, the energy differences between
$m=\pm1$ and $M=\pm 3/2$ are comparable to $eV/2$. All $m$ and $M$ levels are within the bias window (levels $M=\pm 1/2$ are only
slightly outside the window), as shown in Fig.~\ref{fig:B0occ}(c). Thus, the transitions from $m=\pm1$ to $M=\pm3/2$ occur with
significant rates, which gives some occupation probabilities at the $M=\pm 3/2$ levels. Then this induces a series of transitions similar
to the case of $V_g=1.93$ and $V=3.4$~mV, and now the transitions between $M=\pm11/2$ and $m=\pm6$ also occur because the energy difference
between the levels is much smaller due to the increased gate voltage. Compare Fig.~\ref{fig:B0occ}(a) to ~\ref{fig:B0occ}(c).
There are small transition rates from $m=0$ to $M=\pm 1/2$. All $m$ and $M$ levels are occupied to some extent. Therefore, high current
is expected near $V=5.3$~mV. Our analysis of the $I$-$V$ curves at the three different gate voltages clearly reveals that the blockade
effect at low bias cannot be lifted by gate voltage.

\subsection{In the presence of $B_z$ field}

Figure~\ref{fig:Bz}(a)-(d) shows calculated $dI/dV$ as a function of $V$ and $V_g$ at several ${\mathbf B}$ fields when the field is
applied along the $z$ axis. As $B_z$ increases, the spin blockade region becomes narrower and the tip of the V-shaped $dI/dV$ peak shifts
toward higher gate voltage. This blockade effect persists until $B_z$ reaches about 5.2~T at $T=0.05$~meV/$k_B$.
With increasing $B_z$ ($> 0$), the energies of the $M, m < 0$ levels become lower, while those of the $M, m > 0$ levels higher,
due to the Zeeman energy [Fig.~\ref{fig:Bz}(e),(f)]. This level shift modifies the transition rates Eqs.~(\ref{eq:tran1}) and
(\ref{eq:tran2}) and the occupation probabilities $P_{q,r}$ of the levels at given bias and gate voltages. Note that in our model
and Fig.~\ref{fig:Bz} (a)-(d), the tunneling parameters are kept independent of charge state, magnetic level, magnetic field, or gate
voltage. Equations~(\ref{eq:tran1}) and (\ref{eq:tran2}) indicate that such a dependence of the tunneling parameters would change
the transition rates and so it leads to changes in the heights of $dI/dV$ peaks, but these changes would occur only for transitions
allowed by the selection rules and the Fermi-Dirac distribution $f(E)$ within the bias window. Therefore, this dependence would
not eliminate the blockade effect at zero
or low bias. For $B_z > 0$ the lowest level in the cationic state is always $M=-11/2$, whereas the lowest level $-m$ ($m > 0$) in the
neutral state changes with $B_z$. From the energy eigenvalues, we find that this $m$ value is given by a round-off integer of
$g\mu_B B_z/(2|D_0|)$, in agreement with our numerical calculation.

As shown in Fig.~\ref{fig:Bz}(b) and (e), at $B_z=2.0$~T, the lowest level in the $N=0$ state is $m=-2$, and the current
starts to flow at $V=2.0$~mV for $V_g^{\star}=2.72$~mV. For any other gate voltage, higher bias than 2.0~mV is needed for current to flow.
The threshold bias is much lower than that at zero magnetic field, i.e. 3.4~mV. At $V_g^{\star}=2.72$~mV the molecule is in the cationic state
since the $M=-11/2$ level has a lower energy than the $m=-2$ level. At zero bias, only the $M=-11/2$ level is occupied and the transition
rate from $M=-11/2$ to $m=-5$ is zero despite a significant rate of the reverse transition, as shown in Fig.~\ref{fig:occ}(a). The former
transition rate vanishes because of zero occupied electrons at the corresponding energy in the electrodes according to the Fermi-Dirac 
distribution function. As bias increases to
$V=2.0$~mV [Figs.~\ref{fig:occ}(b) and \ref{fig:Bz}(e)], the levels $M=-11/2, -9/2$ and
$m=-4, -3, -2, -1, 0, 1$ enter the bias window and the levels $m=-5, 2$ are slightly outside the bias window. There exists some transition
rate from $M=-11/2$ to $m=-5$, which makes the $m=-5$ level occupied. This triggers ensuing transitions such as
$|m=-5\rangle \leftrightarrow |M=-9/2\rangle \leftrightarrow |m=-4\rangle$ [Figs.~\ref{fig:occ}(b) and~\ref{fig:Bz}(e)]. These transitions
induce significant occupation probabilities at levels $m=-5, -4$ and $M=-9/2$. Then the $M=-7/2$ level is now within the bias window with
respect to the $m=-4$ level, and a transition between $m=-4$ and $M=-7/2$ is now allowed. This gives rise to a substantial occupation
probability at the $M=-7/2$ level. Then a transition between $M=-7/2$ and $m=-3$ is also allowed within the bias window, producing a high
occupation probability at the $m=-3$ level. This whole process contributes to the high current at $V=2.0$~mV. The key step to lift the spin
blockade effect is to produce a substantial occupation probability at the $m=-5$ level via a sizeable transition rate to the $m=-5$ level
from the zero-bias occupied level, $M=-11/2$. As bias increases further to $3.7$~mV, the bias window additionally includes the $m=-6$ level.
Then there is a substantial rate of transition from $M=-11/2$ to $m=-6$ and the transition rate from $M=-11/2$ to $m=-5$ increases.
The increased transition rates trigger transitions similar to the lower-bias case. With respect to the $m=-3$ level, now the $M=-5/2$
level enters the bias window. Thus, there exist a transition between $M=-5/2$ and $m=-3$, and consequently the $M=-5/2$ level is occupied.
A transition from $M=-5/2$ to $m=-2$ now makes the $m=-2$ level occupied, and its reverse transition is also possible. All of these
transitions contribute to the strong $dI/dV$ peak near $V=3.7$~mV, as shown in Fig.~\ref{fig:Bz}(b) and (e). The $V_g^{\star}$ value
differs from the CDP, $V_g=3.03$~mV.

When $B_z$ reaches 4.0~T, $m=-3$ is the lowest level in the $N=0$ state, and the current begins to flow at $V=1.0$~mV for $V_g^{\star}=3.37$~mV [Fig.~\ref{fig:Bz}(c) and (f)]. The threshold bias is much more reduced than the case of $B_z=2.0$~T. At $V_g^{\star}=3.37$~mV the molecule is
in the cationic state. At zero bias, only the $M=-11/2$ level is occupied and the transition rate from $M=-11/2$ to $m=-5$ vanishes despite a significant rate of the reverse transition, due to the same reason as in the case of $B_z=2.0$~T. Thus, the $m=-5$ level is unoccupied and
current cannot flow. For $V=1.0$~mV, the transition from $M=-11/2$ to $m=-5$ occurs and so the $m=-5$ level is occupied [Fig.~\ref{fig:occ}(c)].
Then the $M=-9/2$ level is within the bias window with respect to the $m=-5$ level. Since the transition rate from $m=-5$ to $M=-9/2$ is
significant, the $M=-9/2$ level is occupied. As a result, the transition between $M=9/2$ and $m=-4$ occurs. There are high occupation
probabilities at the $M=-11/2$ and $m=-4$ levels. These series of transitions
$|M=-11/2\rangle \leftrightarrow |m=-5\rangle \leftrightarrow |M=-9/2\rangle \leftrightarrow |m=-4\rangle$
contribute to the high current, as indicated with thick arrows in Fig.~\ref{fig:Bz}(f). The largest contribution to the current arises
from the transition between $m=-5$ and $M=-9/2$.

As $B_z$ increases even further, the $m=-5$ level becomes the lowest in the $N=0$ state. This occurs at $B_z^{\star}=5.78$~T based on
$4.5 = g\mu_B B_z/(2|D_0|)$. Then at the CDP $V_g$ where the $M=-11/2$ and $m=-5$ levels are degenerate, the spin blockade is lifted.
However, $B_z^{\star}$ is slightly higher than the threshold field ($\sim$5.2~T) obtained from our calculated stability diagrams
[Fig.~\ref{fig:Bz}(d)]. This difference can be explained by thermal broadening.
At $B_z=5.2$~T the lowest level in the $N=0$ state is $m=-4$ and the corresponding CDP is 4.02~mV. At this CDP, the energy difference
between $M=-11/2$ and $m=-5$ is about 0.068~meV [Fig.~\ref{fig:occ}(d)]. Considering the thermal energy ($k_B T=0.05$~meV), the $m=-5$
level has some occupancy even at $V=0$. Thus, most of the current at zero bias arises from the transition between $M=-11/2$ and $m=-5$.
Note that in this case the magnitude of the zero-bias current is one order of magnitude smaller than the magnitude of the current at
the first $dI/dV$ peak for lower $B_z$ fields.

We also examine the dependence of the threshold $B_z$ field on $D_0$ and $D_{-1}$. We find that the threshold field depends only on $D_0$,
as long as the zero-field splitting determined by $D_{-1}$ is greater than the thermal energy. The threshold $B_z$ field linearly increases
with increasing $|D_0|$.

\subsection{In the presence of $B_x$ field}

\begin{figure*}
    \includegraphics[width=0.325\textwidth]{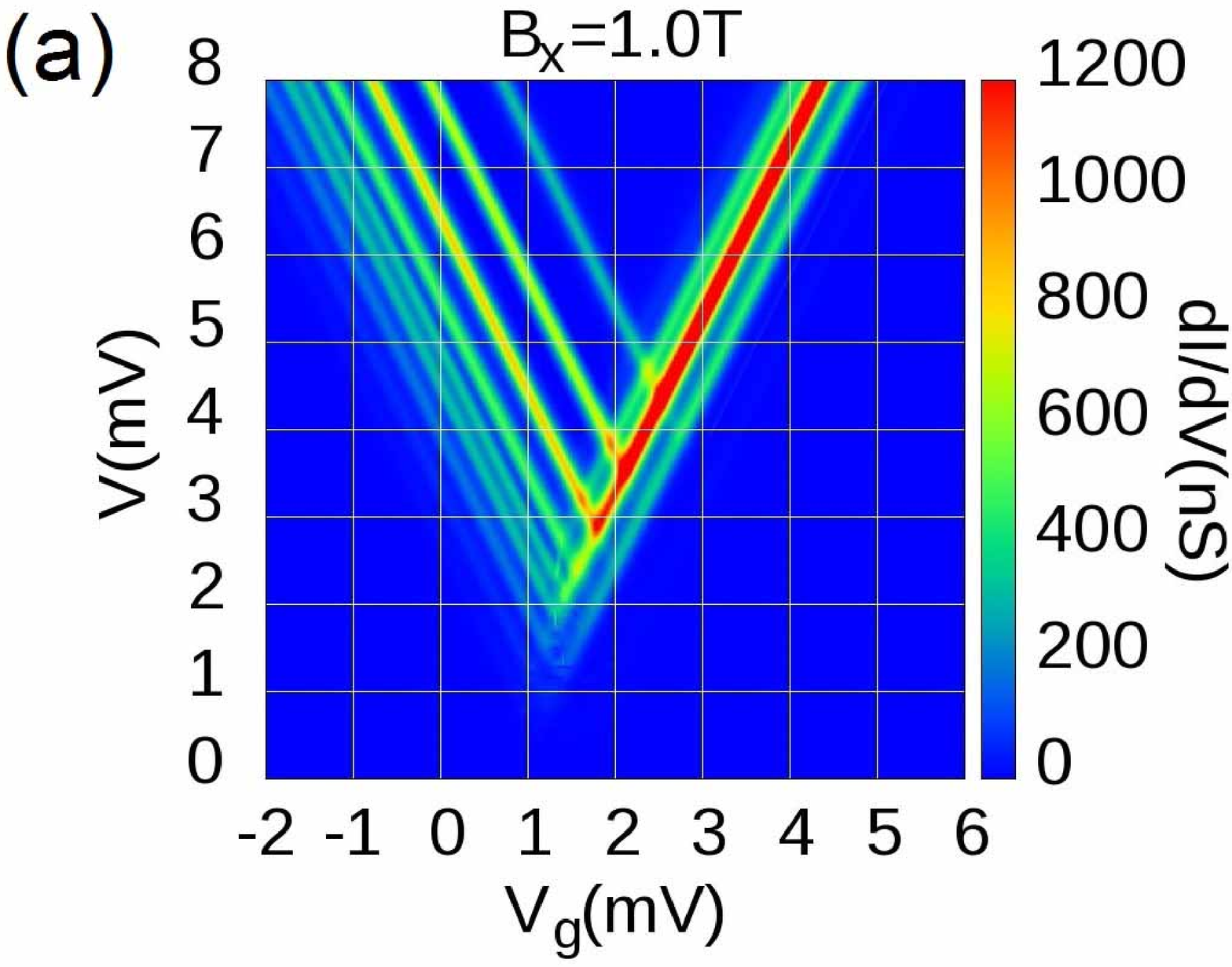}
    \includegraphics[width=0.325\textwidth]{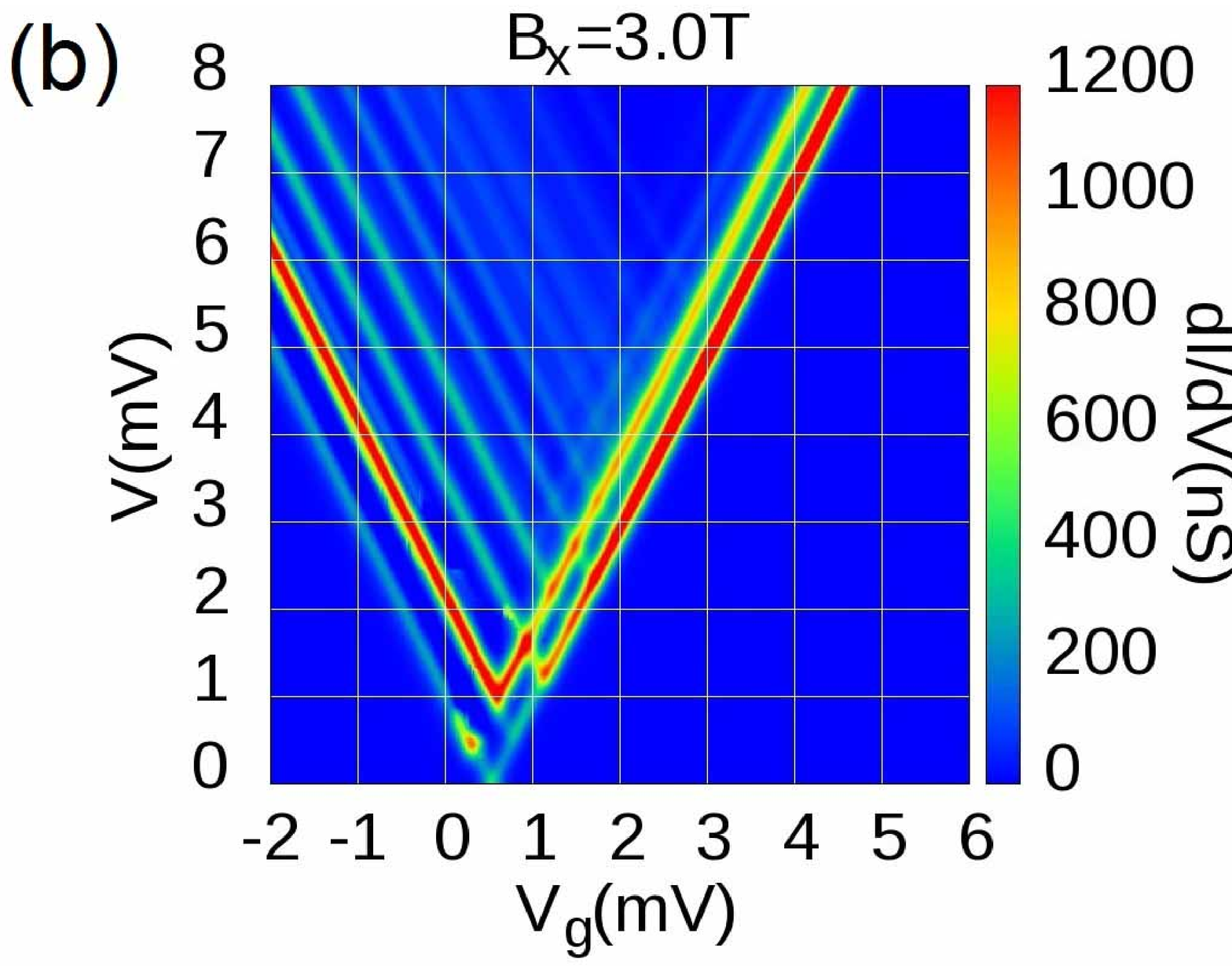}
    \vspace{0.2truecm}
    \includegraphics[width=0.3\textwidth]{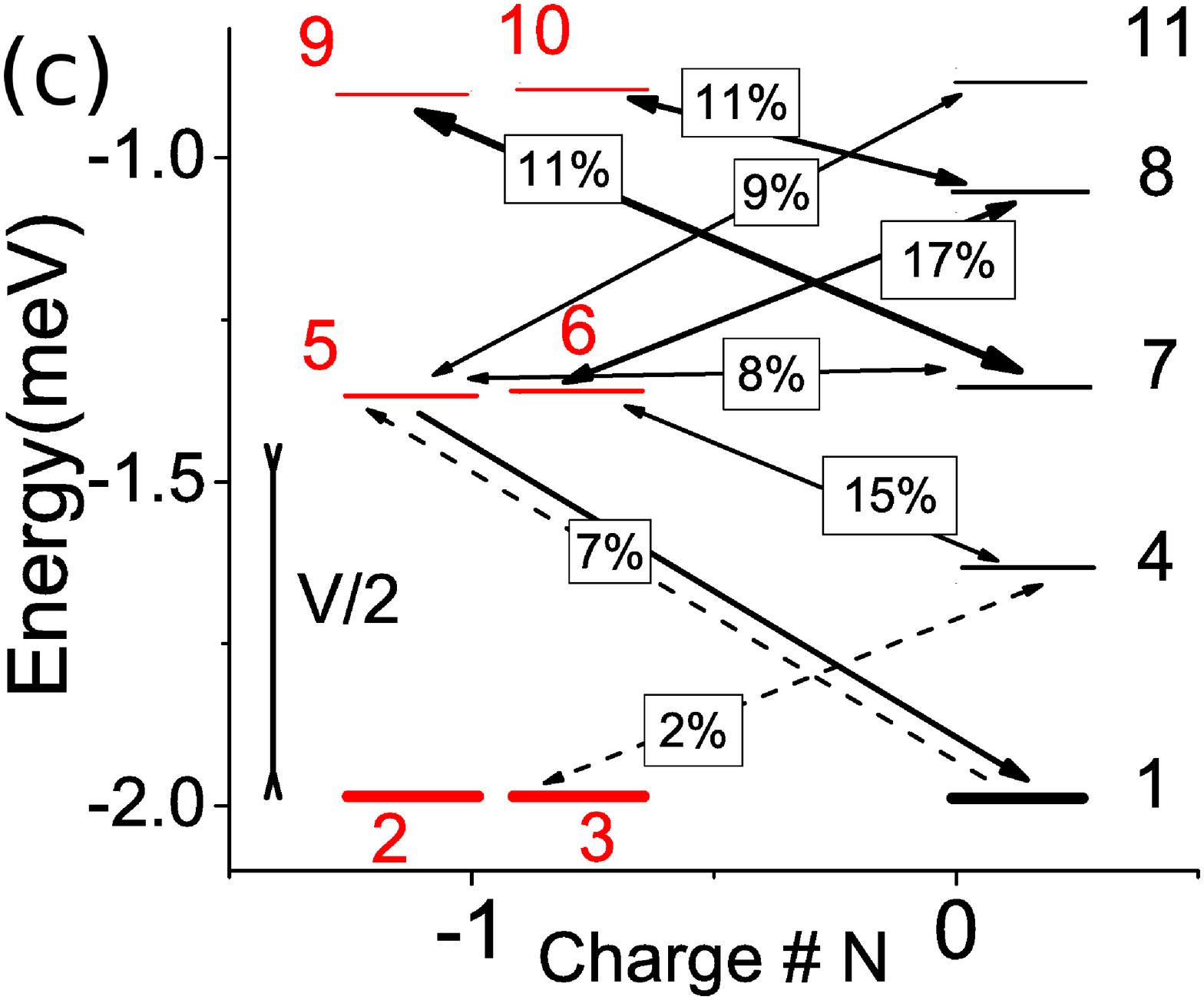}
\caption{(Color online) Calculated $dI/dV$ as a function of $V$ and $V_g$ for (a) $B_x=1.0$ and (b) 3.0~T at $T=0.05$~meV/$k_B$.
(c) Calculated eleven lowest energy levels in the $N=-1$ and $N=0$ states for $B_x=1.0$~T at $V_g=1.40$~mV with their occupation
probabilities, transition rates, and percentage current contributions to the first $dI/dV$ peak at $V=1.1$~mV. Lines, arrows, and
boxed numbers have the same meanings as those in Fig.~\ref{fig:occ}. Levels 1, 4, 7, 8, and 11 are for $N=0$, and the rest of the
levels are for $N=-1$. Here only transition rates contributing to significant current are shown. }
\label{fig:Bx}
\end{figure*}


Figure~\ref{fig:Bx}(a) and (b) shows calculated stability diagrams when ${\mathbf B}$ is along the $x$ axis. As $B_x$ increases, the spin
blockade region becomes narrower and the tip of the V-shaped $dI/dV$ peak shifts toward lower gate voltage. Similarly to the case of the
$B_z$ field, the spin blockade effect is still observed until $B_x$ increases to about 3.0~T at $T=0.05$~meV/$k_B$. The threshold $B_x$ field
is lower than the threshold $B_z$ field. With $B_x$ the molecular eigenstates for each charge state are linear combinations of different
$M$ (or $m$) levels. For example, Fig.~\ref{fig:Bx}(c) shows eleven lowest energy levels at $B_x=1.0$~T for the CDP $V_g=1.40$~mV. The
lowest level for $N=0$ (state 1) is
$-0.07 |m=-3 \rangle + 0.23 |m=-2 \rangle -0.49 |m=-1 \rangle + 0.63 |m=0 \rangle - 0.49 |m=1 \rangle + 0.23 |m=2 \rangle - 0.07 |m=3 \rangle$,
while the two lowest levels for $N=-1$ (states 2 and 3) are
$0.68 |M=-11/2 \rangle - 0.20 |M=-9/2 \rangle + 0.05 |M=-7/2 \rangle \mp 0.05 |M=7/2 \rangle
\pm 0.20 |M=9/2 \rangle \mp 0.68 |M=11/2 \rangle$. At zero bias, the selection rules allow transitions between state 1 and state 2 or 3
due to small contributions of $|m=\pm3 \rangle$ to state 1 and of $|M=\pm7/2 \rangle$ to states 2 and 3. However, their contributions
to the current are negligible at zero bias because of the small coefficients in the eigenstates. There are no other contributions to the
current at zero bias. As $V$ increases, higher levels are accessible and occupied. At $V=1.10$~mV, the excited level, state 4, for $N=0$,
has substantial contributions from $|m=\pm 3 \rangle$ and small additional contributions from $|m=\pm 4 \rangle$, while the excited levels
for $N=-1$ (states 5 and 6) have large contributions from $M=\pm 7/2$ and substantial (small) additional contributions from
$M=\pm 5/2$ ($M=\pm 3/2$). This gives rise to a sizeable rate of transition from state 3 to state 4, which makes state 4 occupied [Fig.~\ref{fig:Bx}(c)]. Then there occur a series of transitions
such as [state 4]$\leftrightarrow$[state 6]$\leftrightarrow$[state 8]$\leftrightarrow$[state 10], within the bias window. Furthermore,
there exists a small rate of transition from state 1 to state 5, which makes state 5 occupied. This leads to two kinds of
transition pathways such as [state 5]$\leftrightarrow$[state 7]$\leftrightarrow$[state 9] and [state 5]$\leftrightarrow$[state 11] among
the eleven lowest energy levels. All of these transitions contribute to the current at $V=1.10$~mV [Fig.~\ref{fig:Bx}(a) and (c)]. As
$B_x$ further increases to 3.0~T, the lowest level in each charge state has contributions from all $M$ or $m$ levels, which allows current
at zero bias [Fig.~\ref{fig:Bx}(b)]. We find that the threshold $B_x$ field depends only on $D_{-1}$, and it decreases as $|D_{-1}|$ decreases.

\section{Conclusion}

We have examined the electron transport properties of the Eu$_{2}$(COT)$_{3}$ molecule weakly coupled to the non-magnetic electrodes
in the sequential tunneling limit. Our calculated $dI/dV$ peaks as a function of $V$ and $V_g$ showed that the current is
strongly suppressed at low bias independently of $V_g$ due to the interplay between the sign reversal of magnetic anisotropy
parameter $D_N$ and the selection rules for the occupied levels. The threshold $B_z$ ($B_x$) field depends only on the
easy-plane (easy-axis) magnetic anisotropy parameter.

\hspace{1.0truecm}

\begin{acknowledgments}
The authors were supported by the U. S. National Science Foundation DMR-1206354.
\end{acknowledgments}


\end{document}